\documentclass[twoside,onecolumn,11pt,technote]{IEEEtran}

\usepackage[ansinew]{inputenc}
\usepackage[dvips]{graphicx}
\usepackage[T1]{fontenc}
\usepackage{here}
\usepackage{amsmath,enumerate,amssymb,hhline}
\newtheorem{definition}{Definition}[section]
\newtheorem{thm}{Theorem}[section]
\newtheorem{proposition}[thm]{Proposition}
\newtheorem{lemma}[thm]{Lemma}
\newtheorem{corollary}[thm]{Corollary}
\newtheorem{exam}{Example}[section]

\newtheorem{remark}{Remark}[section]

\begin{document}
\newcommand{\La}{\mathbf{L}}
\newcommand{\h}{{\bf h}}
\newcommand{\Z}{\mathbb{Z}}
\newcommand{\R}{\mathbb{R}}
\newcommand{\C}{\mathbb{C}}
\newcommand{\F}{\mathbb{F}}
\newcommand{\HH}{{\bf H}}
\newcommand{\OO}{{\cal O}}
\newcommand{\G}{{\cal G}}
\newcommand{\I}{{\cal I}}
\newcommand{\A}{{\cal A}}
\newcommand{\E}{{\cal E}}
\newcommand{\Q}{\mathbb{Q}}
\newcommand{\separ}{\,\vert\,}
\newcommand{\abs}[1]{\vert #1 \vert}

\title{Maximal Orders in the Design of Dense Space-Time Lattice Codes}

  \author{Camilla Hollanti, Jyrki Lahtonen, {\it Member IEEE}, and Hsiao-feng (Francis) Lu \thanks{C. Hollanti is with the  Laboratory of Discrete Mathematics for Information Technology, Turku Centre for Computer Science, Joukahaisenkatu 3-5 B, FIN-20520 Turku, Finland.}
\thanks{C. Hollanti \& J. Lahtonen are with the Department of Mathematics,  FIN-20014 University of Turku, Finland.}
\thanks{E-mails: \{cajoho, lahtonen\}@utu.fi}
\thanks{H.-f. Lu is with the Department of Communication Engineering,
National Chung-Cheng University, Chia-yi, Taiwan}
\thanks{E-mail: francis@ccu.edu.tw}}

  \maketitle

  \begin{abstract}
We construct explicit rate-one, full-diversity, geometrically dense matrix lattices with large, non-vanishing determinants (NVD) for four transmit antenna multiple-input single-output (MISO) space-time (ST) applications. The constructions
are based on the theory of rings of algebraic integers and related
subrings of the Hamiltonian quaternions and can be extended to a larger number of Tx antennas.
The usage of ideals guarantees a non-vanishing  determinant larger than one and an easy way to present the
exact proofs for the minimum determinants. The idea of finding denser sublattices within a given division algebra
is then generalized to a multiple-input multiple-output (MIMO) case with an arbitrary number of Tx antennas
by using the theory of cyclic division algebras (CDA) and maximal orders. It is also shown that the explicit
constructions in this paper all have a simple decoding method based on sphere decoding. Related to the decoding
complexity, the notion of sensitivity is introduced, and experimental evidence indicating a connection between
sensitivity, decoding complexity and performance is provided.
Simulations in a quasi-static Rayleigh fading channel show that our dense quaternionic constructions outperform
both the earlier rectangular lattices and the rotated ABBA lattice as well as the DAST lattice. We also show that our quaternionic lattice is better than the DAST lattice in terms of the diversity-multiplexing gain tradeoff.
  \end{abstract}


  \section{Introduction and background}
\label{intro}
Multiple-antenna wireless communication promises very high data rates, in particular when we have perfect channel state information (CSI) available at the receiver. In \cite{GFBK} the design criteria for such systems were developed and further on the evolution of ST codes took  two directions: trellis codes and block codes. Our work concentrates on the latter branch.

The very first ST block code for two transmit antennas was the  {\it Alamouti code} \cite{Alam} representing
multiplication in the ring of quaternions. As the quaternions form a division algebra,
such matrices must be invertible, i.e. the resulting STBC meets
the rank criterion. Matrix representations of other division algebras
have been proposed as STBCs at least in \cite{HHL}-\cite{EKPKL},
 and (though without explicitly saying so) \cite{WX}. The most recent work \cite{SRS}-\cite{WX}
has concentrated on adding multiplexing gain, i.e. multiple
input-multiple output (MIMO)  applications, and/or combining it
with a good minimum determinant. In this work, we do not
specifically seek any multiplexing gains, but want to improve upon
e.g. the diagonal algebraic space time (DAST) lattices introduced in \cite{DAB} by using
non-commutative division algebras.  Other efforts to improve the
DAST lattices and ideas alike can be found in
\cite{DGB}-\cite{DV}.

The  main contributions  of this work are:
\vspace*{4pt}

\begin{itemize}
\item We give energy efficient
MISO lattice codes with simple decoding that win over e.g. the
rotated ABBA \cite{TBH} and the DAST lattice codes in terms of the block error rate (BLER)
performance.
\item It is shown that by using a non-rectangular lattice one can gain major
energy savings without significant increasement in decoding
complexity. The usage of ideals moreover guarantees a
non-vanishing  determinant $> 1$ and an easy way to present the
exact proofs for the minimum determinants.
\item In addition to the
explicit MISO constructions, we present a general method for
finding dense sublattices within a given CDA in a MIMO setting.
This is tempting as it has been shown in \cite{EKPKL} that CDA-based square ST
codes with NVD achieve the diversity-multiplexing gain 
tradeoff (DMT) introduced in \cite{ZT}. When a CDA is chosen the next step is to choose a corresponding lattice or, what amounts to the same thing, choose an order within the algebra. Most authors, among which e.g. \cite{BORV}, \cite{EKPKL}, and \cite{WX}, have gone with the so-called natural order (see Section \ref{orders}, Example \ref{golden code}). In a CDA based construction, the density of a sublattice  is lumped together with the concept of   maximality of an order. The idea is that one can, on some occasions, use several cosets of the natural order without sacrificing anything in terms of the minimum determinant. So the study of maximal orders is easily motivated by an analogy from the theory of error correcting codes: why one would use a particular code of a given minimum distance and length, if a larger code with  the same parameters is available.
\item Furthermore, related to the decoding complexity, the notion of sensitivity is introduced for the first time, and evidence of its practical appearance is provided. Also the DMT behavior of our codes will be given.
\end{itemize}
\vspace*{4pt}

At first, we are interested in the coherent MISO case with perfect CSI available at the receiver. The received signal $\mathbf{y}\in\C^n$ has the form $$\mathbf{y}=\mathbf{h} X+\mathbf{n},$$
where $X\in \C^{m\times n}$ is the transmitted codeword drawn from a ST code
$\mathcal{C}$, $\mathbf{h}\in\C^m$ is the Rayleigh fading channel
response and the components of the noise vector $\mathbf{n}\in \C^n$ are
i.i.d. complex Gaussian random variables.

A   {\it lattice} is  a discrete finitely generated free
abelian subgroup
of a real or complex finite
dimensional vector space $V$, also called the ambient space. Thus, if $L$ is a $k$-dimensional lattice, there exists a finite set of vectors
${\cal B}=\{\mathbf{b}_1,\mathbf{b}_2,\ldots,\mathbf{b}_k\}\subset V$
such that ${\cal B}$ is linearly independent over the integers
and that
$$
L=\{\sum_{i=1}^k z_i \mathbf{b}_i\mid z_i\in\Z, \mathbf{b_i}\in V\ \textrm{for all } i=1,2,\ldots,k\}.
$$
In the space-time setting a natural ambient space is the space $\C^{n\times n}$
of complex $n\times n$ matrices. When a code is a subset of a lattice $L$
in this ambient space, the
{\it rank criterion} \cite{TSC} states that any non-zero
matrix in $L$ must be invertible. This follows from the fact that the difference of any
two matrices from $L$ is again in $L$.

 The receiver and the decoder, however, (recall that we work in the MISO setting)
observe vector lattices instead of matrix lattices.
When the channel state is $\mathbf{h}$, the receiver expects to see the lattice
$\mathbf{h}L$. If $\mathbf{h}\neq0$ and $L$ meets the rank criterion, then $\mathbf{h} L$
is, indeed, a free abelian group of the same rank as $L$.
However, it is well possible that
$\mathbf{h} L$ is not a lattice, as its generators may be linearly dependent over the reals
--- the lattice is said to {\it collapse}, whenever this happens.

From the pairwise error probability (PEP) point of view \cite{TSC}, the performance of a space-time code is dependent
on two parameters: {\it diversity gain} and {\it coding gain}.
Diversity gain is the minimum of the rank of the difference matrix
$X-X'$ taken over all distinct code matrices
$X,X'\in\mathcal{C}$, also called the {\it rank} of the
code $\mathcal{C}$. When $\mathcal{C}$ is full-rank, the coding
gain is proportional to the determinant of the matrix
$(X-X')(X-X')^H$, where $X^H$ denotes the transpose conjugate of the matrix $X$. The minimum of this determinant
taken over all distinct code matrices is called the {\it minimum
determinant} of the code $\mathcal{C}$ and denoted by $\delta_{\mathcal{C}}$. If $\delta_{\mathcal{C}}$ is bounded
away from zero even in the limit  as SNR $\rightarrow\infty$, the ST
code is said to have the {\it non-vanishing determinant}
property \cite{BRV}. As mentioned above, for non-zero square matrices being full-rank coincides with being invertible.


The  {\it data rate} $R$  in symbols per channel use  is given by
$$ R=\frac{1}{n}\textrm{log}_{|S|}(|\mathcal{C}|), $$ where $|S|$
and $|\mathcal{C}|$ are the sizes of the symbol set and  code
respectively. This is not to be confused with the {\it rate of a
code design} (shortly, {\it code rate}) defined as the ratio of the number of transmitted
information symbols to the decoding delay (equivalently, block
length) of these symbols at the receiver for any given number of
transmit antennas  using any complex signal constellations. If this ratio is equal to the delay, the code is said to have {\it full rate}.

The correspondence is organized as follows: basic  definitions of
algebraic number theory and explicit MISO lattice constructions
are provided in Section \ref{lattices}.  As a (MIMO)
generalization for the idea of finding denser lattices within a
given division algebra, the theory of cyclic algebras and maximal
orders is briefly introduced in Section \ref{cyclic algebras}.  In
Section \ref{decoding}, we consider the decoding of the nested
sequence of quaternionic lattices from Section \ref{lattices}.
A variety of results on decoding complexity  is   established in
Section \ref{decoding}, where also the notion of sensitivity is taken into account.
Simulation results are discussed in
Section \ref{simulation} along with energy considerations. Finally in Section \ref{DMT}, the DMT analysis of the proposed codes will be given.

This work has  been partly  published in a conference, see
\cite{HHL} and \cite{HHL2}. For more background we refer to
\cite{TSC}-\cite{OV}.

\section{Rings of algebraic numbers, quaternions and lattice constructions}
\label{lattices}
We shall denote the sets of integers, rationals, reals, and complex numbers by $\Z,\ \Q,\ \R$, and $\C$ respectively.

Let us  recall the set
$$\mathbb{H}=\{a_1+a_2i+a_3j+a_4k \mid a_t\in\R\ \forall t\},$$
where $i^2=j^2=k^2=-1,\ ij=k$,
 as  the ring of {\it Hamiltonian quaternions}. Note that
 $\mathbb{H}\simeq\mathbb{C}\oplus\mathbb{C} j$,
 when the imaginary unit is identified with $i$. A special interest lies on the  subsets
$$\mathbb{H}_{\mathcal{L}}=\{a_1+a_2i+a_3j+a_4k \mid a_t\in\Z\
\forall t\}\subseteq \mathbb{H} \ \textrm{and}$$
$$\mathbb{H}_{\mathcal{H}}=\{a_1\rho +a_2i+a_3j+a_4k \mid
a_t\in\Z\ \forall t,\ \rho=\textrm{½}(1+i+j+k)\} \subseteq
\mathbb{H}$$ called the {\it Lipschitz'} and {\it Hurwitz'
integral quaternions} respectively.

We shall use extension rings of the Gaussian integers $$\G=\{a+bi
\mid a,b\in\Z\}$$ inside a given division algebra. It would be
easy to adapt the construction to use the slightly denser
hexagonal ring of the Eisensteinian integers $$\E=\{a+b\omega \mid
a,b\in\Z\},$$ where $\omega^3=1$, as a basic alphabet. However, the
Gaussian integers nicely fit with the popular 16-QAM and QPSK
alphabets. Natural examples of such rings are the rings of
algebraic integers inside an extension field  of  the quotient
fields of $\G$, as well as their counterparts inside the
quaternions. To that end we need division algebras $\A$ that are
also 4-dimensional vectors spaces over the field $\Q(i)$.

\subsection{Base lattice constructions}
\label{base lattice} Let now $\zeta=e^{\pi i/8}$ (resp.
$\xi=e^{\pi i/4}=(1+i)/\sqrt2$) be a primitive $16^{th}$ (resp.
$8^{th}$) root of unity. Our main examples of suitable division
algebras are the number field $$\mathbf{L}=\Q(\zeta),$$ and the following
subskewfield  $$\HH=\Q(\xi)\oplus
j\Q(\xi)\subseteq\mathbb{H}$$ of the Hamiltonian quaternions. Note that as $zj=jz^*$ for all complex numbers $z$, and
as the field $\Q(\xi)$ is stable under the usual complex
conjugation $(^*)$, the set $\HH$ is, indeed, a subskewfield of
the quaternions.

As always, multiplication (from the left) by a non-zero element of
a division algebra $\A$ is an invertible $\Q(i)$-linear mapping
(with $\Q(i)$ acting from the right). Therefore its matrix with
respect to a chosen $\Q(i)$-basis ${\cal B}$ of $\A$ is also
invertible. Our example division algebras $\mathbf{L}$ and $\HH$ have the
sets ${\cal B}_L=\{1,\zeta,\zeta^2,\zeta^3\}$ and ${\cal
B}_H=\{1,\xi, j, j\xi\}$ as natural $\Q(i)$-bases. Thus we
immediately arrive at the following matrix representations of our
division algebras. \vspace*{4pt}
\begin{proposition}
\label{algebras}
Let the variables $c_1,c_2,c_3,c_4$ range over all the elements of $\Q(i)$.
The division algebras $\bf{L}$ and $\HH$ can be identified via
an isomorphism $\phi$ with
the following rings of matrices
$$
\mathbf{L}=
\left\{M_L=M_L(c_1,c_2,c_3,c_4)=
\left(
\begin{array}{rrrr}
c_1&ic_4&ic_3&ic_2\\
c_2&c_1&ic_4&ic_3\\
c_3&c_2&c_1&ic_4\\
c_4&c_3&c_2&c_1\\
\end{array}\right)\right\}
$$
and
$$
\HH=\left\{M=M(c_1,c_2,c_3,c_4)=
\left(
\begin{array}{rrrr}
c_1&ic_2&-c_3^*&-c_4^*\\
c_2&c_1&ic_4^*&-c_3^*\\
c_3&ic_4&c_1^*&c_2^*\\
c_4&c_3&-ic_2^*&c_1^*\\
\end{array}\right)\right\}.
$$ The isomorphism $\phi$ from $\bf{L}$ into the matrix ring is
determined by $\Q(i)$-linearity and the fact that $\zeta$
corresponds to the choice $c_2=1, c_1=c_3=c_4=0$. The isomorphism
$\phi$ from $\HH$ into the matrix ring is determined by
$\Q(i)$-linearity and the facts that $\xi$ corresponds to the
choice $c_2=1$, $c_1=c_3=c_4=0$, and $j$ corresponds to the choice
$c_3=1$, $c_1=c_2=c_4=0$. In particular, the determinants of these
matrices are non-zero whenever at least one of the coefficients
$c_1,c_2,c_3,c_4$ is non-zero.\hfill  \QED
\end{proposition}
\vspace*{4pt}

In order to get ST lattices and useful bounds for
the minimum determinant, we need to identify suitable subrings $S$
of these two algebras. Actually, we would like these rings to be
free right $\G$-modules of rank 4. This is due to the fact that
then the determinants of the matrices of Proposition
\ref{algebras} that belong to the subring $\phi(S)$ must be
elements of the ring $\G$. We repeat the well-known reason for
this for the sake of completeness: the determinant of the matrix
representing the multiplication by a fixed element $x\in S$ does
not depend on the choice of the basis ${\cal B}$ and thus we may
assume that it is  a $\G$-module basis. However, in that case
$x{\cal B}\subseteq S$, so the matrix will have entries in $\G$ as
all the elements of $S$ are $\G$-linear combinations of ${\cal
B}$. The claim follows.

\newcommand{\lip}{{\cal L}}
\newcommand{\hur}{{\cal H}}

In the case of the field $\bf{L}$ we are only interested in its ring of
integers $\OO_L=\Z\lbrack\zeta\rbrack$ that is a free $\G$-module
with  the basis ${\cal B}_L$. In this case the ring $\phi(\OO_L)$
consists of those matrices of $\bf{L}$ that have all the coefficients
$c_1,c_2,c_3,c_4\in\G$. Similarly, the $\G$-module
$$\lip=\G\oplus\xi\G\oplus j\G\oplus j\xi\G$$ spanned by our
earlier basis ${\cal B}_H$ is a ring  of the required type. We
call this
 the ring of {\it Lipschitz' integers of $\HH$}. Again $\phi(\lip)$
consists of those matrices of $\HH$ that have all the coefficients
$c_1,c_2,c_3,c_4\in\G$. While $\OO_L$ is known to be maximal among
the rings satisfying our requirements, the same is not true about
$\lip$. The ring $\mathbb{H}_{\hur}$ also has an extension of the
prescribed type inside $\HH$, called the ring of {\it Hurwitz'
integers of $\HH$}. This ring, denoted by
$$\hur=\rho\G\oplus\rho\xi\G\oplus j\G\oplus j\xi\G,$$ is the
right $\G$-module generated by the basis ${\cal
B}_{Hur}=\{\rho,\rho\xi,j,j\xi\}$, where again $\rho=(1+i+j+k)/2$.
The fact that $\hur$ is a subring can easily be verified by
straightforward computations, e.g. $\xi\rho=\rho\xi-j\xi$. For
future use we express the ring $\hur$ in terms of the basis ${\cal
B}_H$ of Proposition \ref{algebras}. It is not difficult to see
that the element $$ q=c_1+\xi c_2 + j c_3 + j\xi c_4\in\HH $$ is
an element of $\hur$, if and only if the coefficients $c_t$
satisfy the requirements $(1+i)c_t\in\G$ for all $t=1,2,3,4$ and
$c_1+c_3,c_2+c_4\in\G$. As the ideal generated by $1+i$ has
index two in $\G$, we see that $\lip$ is an additive, index four subgroup in $\hur$. We summarize these findings in Proposition
\ref{rings}. The bound on the minimum determinant is a consequence
of the fact that all the elements of $\G$ have a norm at least one.
\vspace*{4pt}
\begin{proposition}
\label{rings}
The following rings of matrices form ST lattices with minimum
determinant equal to one.
$$
L_1=
\left\{
M_L(c_1,c_2,c_3,c_4) \mid  c_1,c_2,c_3,c_4\in\G\right\},
$$
$$
L_2=\left\{M(c_1,c_2,c_3,c_4)
\mid c_1,c_2,c_3,c_4\in\G\right\},
$$
$$
L_3=\left\{M(c_1,c_2,c_3,c_4)
\mid c_1,c_2,c_3,c_4\in\frac{1+i}{2}\G,\ c_1+c_3\in\G,c_2+c_4\in\G\right\}.
$$\hfill\QED
\end{proposition}
\vspace*{4pt}

\begin{remark} The lattice $L_1$ is quite similar to the DAST lattice in the sense that all
of its matrices can be simultaneously diagonalized. See more details in Section \ref{complexity}.
The lattice $L_2$, for its part, is a more developed
case from the so-called {\it quasi-orthogonal} STBC suggested e.g. in \cite{J}.
The matrix $M(c_1,c_2,c_3,c_4)$ of Proposition \ref{algebras}   can also be found as an example in the landmark paper \cite{SRS},
but no optimization has been done there by using, for example, ideals as we shall do here.
\end{remark}
\vspace*{4pt} A drawback shared by the lattices $L_1$ and $L_2$ is
that in the ambient space of the transmitter they are isometric to
the rectangular lattice $\Z^8$. The rectangular shape does carry
the advantage that the sets of information carrying coefficients
of the basis matrices are simple and all identical which is useful
in e.g. sphere decoding. But, on the other hand, this shape is
very wasteful in terms of transmission power. Geometrically denser
sublattices of $\Z^8$, e.g. the checkerboard lattice $$
D_8=\left\{(x_1,...,x_8)\in\Z^8\ \bigg|\  \sum_{i=1}^8 x_i\equiv
0\ (\textrm{mod }2)\right\} $$
 and
the diamond lattice
$$
E_8=\left\{(x_1,...,x_8)\in\Z^8\ \bigg|\  x_i\equiv x_j\ (\textrm{mod }2),\ \sum_{i=1}^8 x_i \equiv 0\ (\textrm{mod }4)\right\},
$$
 are well-known (cf. e.g. \cite{CS}). However, we
must be careful in picking the copies of the sublattices, as it is the
minimum determinant we want to keep an eye on (see Remark \ref{determinant vs distance}).

\subsection{Dense sublattices inside the base lattice $L_2$}
\label{dense sub}

As our earlier simulations \cite{HHL},\cite{HHL2} have shown that
$L_2$ outperforms $L_1$, we concentrate on finding good
sublattices of $L_2$. The units of the ring $L_2$ are exactly the
non-zero matrices whose determinants have the minimal absolute
value of one. Thus a natural way to find a sublattice with a
better minimum determinant is to take the lattice $\phi(\I)$,
where $\I\subset S$ is a proper ideal. This idea has appeared at least in
\cite{HHL}, \cite{HHL2}, and \cite{BRV}. Even earlier, ideals of rings of
algebraic integers were used in \cite{BVRB} to produce dense
lattices. Let us first record the following simple fact.
\vspace*{4pt}
\begin{lemma}
\label{detineq} Let $A$ and $B$ be diagonalizable complex square
matrices of the same size. Assume that they commute and that their
eigenvalues are all real and non-negative. Then $$ \det{(A+B)}\ge
\det{A}+\det{B} $$ with a strict inequality if both $A$ and $B$
are invertible.
\end{lemma}
\vspace*{4pt}
\begin{proof}
As $A$ and $B$ commute, they can be simultaneously diagonalized.
Hence, we can reduce the claim to the case of diagonal matrices
with non-negative real entries. In that case the claim is
obvious.
\end{proof}
\vspace*{4pt}

In Proposition \ref{D8} we give a construction isometric to the checkerboard lattice $D_8$
\begin{proposition}
\label{D8} Let $\I$ be the prime ideal of the ring $\G$ generated
by $1+i$. Define $$ \I_{\lip}=\{(c_1+\xi c_2)+j(c_3+\xi
c_4)\in\lip \mid c_1+c_2+c_3+c_4\in{\cal I}\}. $$ Then $\I_{\lip}$
is an ideal of index two in $\lip$. The corresponding lattice $$
L_4=\{M(c_1,c_2,c_3,c_4)\in L_2 \mid c_1+c_2+c_3+c_4\in{\cal I}\}
$$ is an index $2$ sublattice in $L_2$. Furthermore, the absolute
value of $\det(MM^H),\ M\in L_4\setminus \{0\}$, is then at least
$4$.
\end{proposition}
\vspace*{4pt}

\begin{proof}
It is straightforward to check that ${\cal I}_\lip$ is stable
under (left or right) multiplication with the quaternions $\xi$ and $j$,
so ${\cal I}_\lip$ is an ideal in $\lip$.

Let us consider a matrix $M\in L_4$ and write it in the block form
$$
M=
\left(
\begin{array}{rr}
A&-B^H\\
B&A^H
\end{array}\right).
$$
We see that
$$
MM^H=
\left(
\begin{array}{cc}
AA^H+BB^H&0\\
0&AA^H+BB^H
\end{array}
\right),
$$
and
$$
AA^H+BB^H=
\left(
\begin{array}{cc}
\alpha&k^*\\
k&\alpha
\end{array}\right),
$$
where $\alpha=\sum_{j=1}^4\abs{c_j}^2$ is a non-negative
integer and $k=-ic_1c_2^*+c_2c_1^*-ic_3c_4^*+c_4c_3^*$
is a Gaussian integer with the property $k^*=ik$.
We are to prove that $\det{MM^H}=\left(\alpha^2-\abs{k}^2\right)^2\ge4$.
Assume first that $c_3=c_4=0,$ i.e. the block $B=0$. Then
$\det(A)$ is the relative norm $$\det(A)=N^{\Q(\xi)}_{\Q(i)}(c_1+\xi c_2),$$
which is a Gaussian integer. As $c_1+\xi c_2$ is a non-zero element of the
ideal $\cal I$, we conclude that $\det(A)$ is a non-zero non-unit.
Therefore $\det(A)\det(A^H)\ge 2$, and the claim follows.

Let us then assume that both $A$ and $B$ are non-zero. Then $\det(A)$
and $\det(B)$ are non-zero Gaussian integers and have a norm at least one.
The matrices $A,A^H,B,B^H$ all commute, so by Lemma \ref{detineq}
we get
$$
\det(MM^H)> \det(AA^H)^2+\det(BB^H)^2\geq 2.
$$
As $\det(MM^H)=\left(\alpha^2-\abs{k}^2\right)^2$ is a square of
a rational integer, it must be at least 4.
\end{proof}
\vspace*{4pt}

\begin{remark}
\label{D8 remark} It is easy to see that in the previous
proposition $a+bi\in{\cal I}$, if and only if $a+b$ is an even integer. Thus
geometrically the matrix lattice $L_4$ is, indeed, isometric to
$D_8$.
\end{remark}
\vspace*{4pt}

We proceed to describe two more interesting sublattices of $L_2$
with even better minimum determinants. To that end we use the ring
$\hur$ (or the lattice $L_3$). The first sublattice is isometric
to the direct sum $D_4\perp D_4$ \cite{CS}
 of two 4-dimensional checkerboard lattices.
\vspace*{4pt}

\begin{proposition}
\label{D4D4} Let again $\cal I$ be the ideal $(1+i)\G$. The
lattice $$ L_5=\left\{M(c_1,c_2,c_3,c_4)\in L_2 \mid
c_1+c_3,c_2+c_4\in{\cal I}\right\}$$ has a minimum determinant
equal to 16. The index of $L_4$ in $L_2$ is $4$.
\end{proposition}
\vspace*{4pt}

\begin{proof}
The coefficients $c_1$ and $c_3$ can be chosen arbitrarily within $\G$. The the ideal $\cal I$ has index $2$ in $\G$, and the coefficients $c_2$ and $c_4$ now must belong to the cosets $c_1+\cal I$ and $c_3+\cal I$ respectively. Whence, the index of $L_5$ in $L_2$ is 4.
The matrices $A$ in the lattice $L_5$ are of the form $A=(1+i)M$,
where $M$ is a matrix in the lattice $L_3$ of Proposition
\ref{rings}. Thus $\det(AA^H)=16 \det(MM^H)$ and the claim follows
from Proposition \ref{rings}.
\end{proof}
\vspace*{4pt} The diamond lattice $E_8$ can be described in terms
of the Gaussian integers as  (cf. \cite{A})
$$E_8=\frac{1}{1+i}\left\{(c_1,c_2,c_3,c_4)\in\G^4 \mid
 c_1+\I=c_t+\I,\ t=2,3,4,\
\sum_{t=1}^4 c_t\in 2\G\right\}.$$ By our identification of quadruples
$(c_1,c_2,c_3,c_4)\in\G^4$ and the elements of $\HH$ it is
straightforward to  verify that $(1+i)E_8$ has
$\{2,(1+i)+(1+i)\xi,(1+i)\xi+(1+i)j, 1+\xi+j+j\xi\}\subseteq\lip$
as a $\G$-basis, whence the set $\{1+i,1+\xi,\xi+j,
\rho+\rho\xi\}\subseteq \hur$ is a $\G$-basis for $E_8$. By
another simple computation we see that $E_8=\hur(1+\xi)$, i.e.
$E_8$ is the left ideal of the ring $\hur$ generated by $1+\xi$.
\vspace*{4pt}

\begin{proposition}
\label{E8}
 The lattice
$$L_6=\left\{M(c_1,c_2,c_3,c_4)\in L_2
\separ
\ c_1+\I= c_t+\I,\ t=2,3,4,\
\sum_{t=1}^4 c_t\in 2\G\right\}$$
is an index 16 sublattice of $L_2$.
Furthermore, the minimum determinant of $L_6$ is $64$.
\end{proposition}
\vspace*{4pt}

\begin{proof}
Let $M_I=M(1,1,0,0)$ be the matrix $\phi(1+\xi)$ under the
isomorphism of Proposition \ref{algebras}. We see that
$\det(M_IM_I^H)=4$. By the preceding discussion any matrix $A$ of
the lattice $L_6$ has the form $A=M M_I (1+i)$, where $M$ is a
matrix in $L_3$. As in the proof of Proposition \ref{D4D4}, we see
that $\det{AA^H}=16\det(M_IM_I^H)\det(MM^H)$. The claim
on the minimum determinant now follows from Proposition \ref{rings}.
We see that the coefficient $c_1$ can be chosen arbitrarily within
$\G$. The coefficients $c_2$ and $c_3$ then must belong to the
coset $c_1+{\cal I}$, and $c_4$ must be chosen such that
$c_1+c_2+c_3+c_4\in 2\G={\cal I}^2$. As ${\cal I}$ has index two
in $\G$, we see that the index of $L_6$ in $L_2$ is 16 as claimed.
\end{proof}
\vspace*{4pt}
\begin{remark}
\label{determinant vs distance}
We have now produced a {\it nested sequence of lattices}
\begin{equation}
\label{nested sequence}
2\Z^8=2L_2\subseteq L_6\subseteq L_5\subseteq L_4\subseteq L_2=\Z^8 (\subseteq L_3).
\end{equation}
We concentrate on the lattices that are sandwiched between $2\Z^8$
and $\Z^8$. It is worthwhile to note that these lattices are in a
bijective correspondence with a binary linear code of length 8 by
projection modulo 2, see Table \ref{binary} above. As it happens, within
this sequence of lattices the minimum Hamming distance of the
binary linear code and the minimum determinant of the lattice are
somewhat related.

\begin{table}
\caption{Lattices from a coding theoretical point of view}
\label{binary}
\begin{center}
\renewcommand{\arraystretch}{1.44}
\begin{tabular}{|c|}
 \hline
 $L_2\leftrightarrow$ The 8-dimensional rectangular grid $\Z^8$
$\leftrightarrow$ no coding\\ $\downarrow$\\ $L_4\leftrightarrow$ The checkerboard
lattice $D_8$ $\leftrightarrow$ overall parity check code of
length $8$\\ $\downarrow$\\ $L_5\leftrightarrow$ The lattice $D_4\perp D_4$
$\leftrightarrow$ two blocks of the overall parity check code of
length $4$\\ $\downarrow$\\ $L_6\leftrightarrow$ The diamond lattice $E_8$
$\leftrightarrow$ extended Hamming-code of length $8$\\ \hline
\end{tabular}
\end{center}
\end{table}

Thereupon it is natural to ask that what if we simply concatenate
the use of $L_2$ with a good binary code (extended over several
$L_2$-blocks, if needed), and be done with it. While the binary
linear codes appearing above are the first ones that come to one's mind,
we want to caution the unwary end-user. Namely, it is possible
that there are high weight units in the ring in question. If such
binary words are included, then the minimum determinant of the
corresponding lattice is equal to $1$, i.e. no coding gain will
take place. E.g. the unit
$(1-\xi^3)/(1-\xi)=1+\xi+\xi^2=(1+i)+\xi$ of the ring $\lip$
corresponds to the matrix $M(1+i,1,0,0)$ of determinant 1, and
thus we must not allow such words of weight 3. If the lattice
$L_1$ were used, the situation would be even worse, as then we
have units like $(1-\zeta^7)/(1-\zeta)$ in the ring $\OO_L$ that
would be mapped to a word of Hamming weight 7. A construction
based on ideals provides a mechanism to avoid this problem caused
by high weight units.
\end{remark}

\section{Cyclic algebras and orders}
\label{cyclic algebras}

In Section \ref{lattices} we produced a nested sequence (\ref{nested sequence}) of
quaternionic lattices with the property that as the lattice gets
denser after rescaling the  increased minimum determinant back to one, the BLER perfomance gets better. As the sequence (\ref{nested sequence}) lies
within a specific division algebra, an obvious question evokes
how to generalize this idea. The theory of cyclic division
algebras and their maximal orders offer us an answer. When
designing square ST matrix lattices for MIMO use, cyclic division algebras are
of utmost interest as it has been shown in \cite{EKPKL} that a
non-vanishing determinant is a sufficient condition for full-rate
CDA based STBC-designs to achieve the upper bound on the optimal DMT, hence proving that the upper bound itself is the optimal
DMT for any number of transmitters and receivers. Given
the number of transmitters $n$, we pick a suitable  cyclic
division algebra of index $n$ (more on this in a forthcoming
paper, see Section \ref{conclusions} and \cite{HLRV}. See also \cite{EKPKL}
). The
matrix representation of the algebra, with some constraints on the
elements, will then correspond to the base lattice, similarly as
did the lattice $L_2$ in Section \ref{lattices}. Now in order to
make the lattice denser, we choose the elements in the matrices
from an order. The natural first choice for an order is the one corresponding to the ring
of algebraic integers of the maximal subfield inside the algebra.
The densest possible sublattice is the one where the elements come
from a maximal order.

All algebras considered here are finite dimensional associative
algebras over a field.

\subsection{Cyclic algebras}

The basic theory of cyclic algebras and their representations  as
matrices are thoroughly considered in [\cite{Jac}, Chapter 8.5] and \cite{SRS}. We are only
going to recapitulate the essential facts here.

In the following, we consider number field extensions $E/F$, where
$F$ denotes the base field. $F^*$ (resp. $E^*$) denotes the set of
non-zero elements of $F$ (resp. $E$). Let $E/F$ be a cyclic field
extension of degree $n$ with the Galois group $Gal(E/F)=\left\langle
\sigma\right\rangle$, where $\sigma$ is the generator of the
cyclic group. Let $\mathcal{A}=(E/F,\sigma,\gamma)$ be the
corresponding cyclic algebra of {\it index} $n$, that is,
\begin{center}
$\mathcal{A}=E\oplus uE\oplus u^2E\oplus\cdots\oplus u^{n-1}E$,
\end{center}
with $u\in\mathcal{A}$ such that $xu=u\sigma(x)$ for all $x\in E$
and $u^n=\gamma\in F^*$. An element
$a=x_0+ux_1+\cdots+u^{n-1}x_{n-1}\in\mathcal{A}$ has the following
representation as a matrix $A=$
\begin{equation}
\label{CA}
\left(
\begin{array}{ccccc}
x_0&\gamma \sigma(x_{n-1})&\gamma \sigma^2(x_{n-2})&\cdots & \gamma \sigma^{n-1}(x_1)\\
x_1&\sigma(x_0)&\gamma\sigma^2(x_{n-1})& &\gamma\sigma^{n-1}(x_2)\\
x_2&\sigma(x_1)&\sigma^2(x_0)& &\gamma\sigma^{n-1}(x_3)\\
\vdots& & & & \vdots\\
x_{n-1}&\sigma(x_{n-2})&\sigma^2(x_{n-3})&\cdots&\sigma^{n-1}(x_0)\\
\end{array}\right).
\end{equation}

Let us compute the third column as an example:
\begin{eqnarray*}
u^2\mapsto au^2&=&x_0u^2+ux_1u^2+\cdots+u^{n-1}x_{n-1}u^2\\
&=&u\sigma(x_0)u+u^2\sigma(x_1)u+\cdots+\gamma\sigma(x_{n-1})u\\
&=&u^2\sigma^2(x_0)+u^3\sigma^2(x_1)+\cdots+u\gamma\sigma^2(x_{n-1}),
\end{eqnarray*}
and hence as the third column we get the vector
$$(\gamma\sigma^2(x_{n-2}),\gamma\sigma^2(x_{n-1}),\sigma^2(x_0),\ldots,\sigma^2(x_{n-3}))^T.$$
\bigbreak Let us denote the ring of algebraic  integers of $E$ by
$\OO_E$. A basic, rate-$n$ MIMO STBC $\mathcal{C}$ is usually
defined as $\mathcal{C}=$
\begin{equation}
\label{cyclic stbc}
\left\{\left(
\begin{array}{cccc}
x_0&\gamma \sigma(x_{n-1})&\cdots & \gamma \sigma^{n-1}(x_1)\\
x_1&\sigma(x_0)& &\gamma\sigma^{n-1}(x_2)\\
x_2&\sigma(x_1)& &\gamma\sigma^{n-1}(x_3)\\
\vdots& & & \vdots\\
x_{n-1}&\sigma(x_{n-2})&\cdots&\sigma^{n-1}(x_0)\\
\end{array}\right)\Bigg|\ x_i\in \OO_E\right\}.
\end{equation}
Further optimization might  be carried out by using e.g. ideals.
If we denote the basis of $E$ over  $\OO_F$ by
$\{1,e_1,...,e_{n-1}\}$, then  the elements $x_i,\ i=0,...,n-1$ in
(\ref{cyclic stbc}) take the form $x_i=\sum_{k=0}^{n-1}f_ke_k$,
where $f_k\in \OO_F$ for all $k=0,...,n-1$. Hence $n$ complex
symbols are transmitted per channel use, i.e. the design has rate $n$. In literature this is often referred to as having a {\it full rate}. \vspace*{4pt}

\begin{definition}
An algebra $\mathcal{A}$ is called {\it simple} if it has no
nontrivial ideals. An $F$-algebra $\mathcal{A}$ is {\it central} if
its center $Z(A)=\{a\in\mathcal{A}|aa'=a'a\ \forall
a'\in\mathcal{A}\}=F$.
\end{definition}
\vspace*{4pt}

\begin{definition}
An ideal $\mathcal{I}$ is called {\it nilpotent} if $\mathcal{I}^k=0$ for some $k\in \Z_+$.
 An algebra $\mathcal{A}$ is {\it semisimple} if it has no
nontrivial nilpotent ideals. Any finite dimensional semisimple algebra over a field is a finite and unique direct sum of simple algebras.
\end{definition}
\vspace*{4pt}

\begin{definition}
The determinant (resp. trace) of the matrix $A$ is called the {\it
reduced norm} (resp. {\it reduced trace}) of an element
$a\in\mathcal{A}$ and is denoted by $nr(a)$ (resp. $tr(a)$).
\end{definition}
\vspace*{4pt}
\begin{remark} The connection with the usual norm map  $N_{A/F}(a)$   (resp. trace map $T_{A/F}(a)$)
 and the reduced norm $nr(a)$ (resp. reduced trace $tr(a)$) of an element $a\in\mathcal{A}$ is $N_{A/F}(a)=(nr(a))^n$
 (resp. $T_{A/F}(a)=n tr(a)$), where $n$ is the degree of $E/F$.
\end{remark}
\vspace*{4pt}

 In Section \ref{lattices} we have attested that the
algebra $\HH$ is a division algebra. The next old result due to A. A. Albert [\cite{AA}, Chapter V.9] provides
us with a condition for when an algebra is indeed a division
algebra. \vspace*{4pt}
\begin{proposition}
The algebra $\mathcal{A}=(E/F,\sigma,\gamma)$ of index $n$ is a division algebra, if and only if  the smallest factor $t\in\Z_+$ of $n$ such that
$\gamma^t$ is the norm of some element in $E^*$, is $n$. \hfill  \QED
\end{proposition}

\subsection{Orders}
 \label{orders}
We are now ready to present some of the basic definitions and results from the theory of maximal orders.
The general theory of maximal orders can be found in \cite{R}.

Let $S$ denote a Noetherian integral domain with a quotient field
$F$, and let $\mathcal{A}$ be a finite dimensional $F$-algebra.
\vspace*{4pt}
\begin{definition}
An $S${\it -order} in the $F$-algebra $\mathcal{A}$ is a subring
$\Lambda$ of $\mathcal{A}$, having the same identity element as
$\mathcal{A}$, and such that $\Lambda$ is a finitely generated
module over $S$ and generates $\mathcal{A}$ as a linear space over
$F$.
\end{definition}
\vspace*{4pt}

As usual, an  $S$-order in $\mathcal{A}$ is
said to be {\it maximal},  if it is not properly contained in any
other $S$-order in $\mathcal{A}$. If the integral closure
$\overline{S}$ of  $S$ in $\mathcal{A}$ happens to be an $S$-order
in $\mathcal{A}$, then $\overline{S}$ is automatically the unique
maximal $S$-order in $\mathcal{A}$.

Let us illustrate the above definition by the following example.
\vspace*{4pt}
\begin{exam}
(a) Orders always exist: If $M$ is a full $S$-lattice in
$\mathcal{A}$, i.e. $FM=\mathcal{A}$, then the {\it left order} of
$M$ defined as $\OO_l(M)=\{x\in\mathcal{A} \mid xM\subseteq M\}$ is an
$S$-order in $\mathcal{A}$. The right order is defined in an
analogous way.

(b) If $\mathcal{A}=\mathcal{M}_n(F)$, the algebra of all $n\times
n$ matrices over $F$, then $\Lambda=\mathcal{M}_n(S)$ is an
$S$-order in $\mathcal{A}$.

(c) Let $a\in\mathcal{A}$ be integral over $S$, that is, $a$ is a
zero of a monic polynomial over $S$. Then  the ring $S[a]$ is an
$S$-order in the $F$-algebra $F[a]$.

(d) Let $S$ be a Dedekind domain, and let $E$ be a finite
separable extension of $F$. Denote by $\overline{S}$ the integral
closure of $S$ in $E$. Then $\overline{S}$ is an $S$-order in $E$.
In particular, taking $S=\Z$, we see that the ring of algebraic
integers  of $E$ is a $\Z$-order in $E$.
\end{exam}
\vspace*{4pt}

Hereafter, $F$ will be an algebraic number field and $S$ a Dedekind ring with $F$ as a field of fractions.
\vspace*{4pt}
\begin{proposition}
\label{ronyaile 2.5} Let
$\mathcal{A}$ be a finite dimensional semisimple algebra over $F$ and
$\Lambda$ be a $\Z$-order in $\mathcal{A}$. Let $\OO_F$ stand for
the ring of algebraic integers of $F$.  Then $\Gamma=\OO_F\Lambda$
is an $\OO_F$-order containing $\Lambda$. As a consequence, a
maximal $\Z$-order in $\mathcal{A}$ is a maximal $\OO_F$-order as
well. \hfill  \QED
\end{proposition}
\vspace*{4pt}

The following proposition provides us with a useful tool for  finding a maximal order within a given algebra.
\vspace*{4pt}
\begin{proposition}
\label{normiehto}
Let $\Lambda$ be an $S$-order in $\mathcal{A}$. For each $a\in\Lambda$ we have
$
nr(a)\in S\textrm{ and}\ tr(a)\in S.
$
\hfill  \QED
\end{proposition}

\vspace*{4pt}
\begin{proposition}
Let $\Gamma$ be a subring of $\mathcal{A}$ containing $S$, such that $F\Gamma=\mathcal{A}$, and suppose that each $a\in\Gamma$ is integral over $S$. Then $\Gamma$ is an $S$-order in $\mathcal{A}$. Conversely, every $S$-order in $\mathcal{A}$ has these properties.
\hfill  \QED
\end{proposition}
\vspace*{4pt}

\begin{corollary}
\label{maxcor}
Every $S$-order in $\mathcal{A}$ is contained in a maximal $S$-order in $\mathcal{A}$. There exists at least one maximal $S$-order in $\mathcal{A}$.
\hfill  \QED
\end{corollary}
\vspace*{4pt}
\begin{remark}
As the previous corollary indicates, a maximal order of an algebra is not necessarily unique.
\end{remark}
\vspace*{4pt}

\begin{remark}
The algebra $\HH$ can also be viewed as a cyclic division algebra.
As it is a subring of the Hamiltonian quaternions, its center
consists of the intersection $\HH\cap\R=\Q(\sqrt2)$. Also
$\Q(\xi)$ is an example of a splitting field of $\HH$. In the
notation above we have an obvious isomorphism
$$ \HH\simeq
(\Q(\xi)/\Q(\sqrt2),\sigma,-1), $$
where $\sigma$ is the usual
complex conjugation.
\end{remark}
\vspace*{4pt}

\begin{remark}
\label{vanishing}
In principle,  the lattices from Section \ref{lattices} could also be used as MIMO codes, but when we  pack $\HH$ in the form of (\ref{CA}), $\delta_{\mathcal{C}}$ becomes vanishing and the DMT cannot be achieved.
\end{remark}

\vspace*{4pt}

One extremely well-performing CDA based code taking advantage of a maximal order  is the celebrated {\it Golden code} \cite{BRV} (also independently found in \cite{YW}) treated in the following example.
\vspace*{4pt}
\begin{exam}
\label{golden code} In any cyclic algebra where the element $\gamma$
 happens to be an algebraic
integer, we have the following {\it natural order}
$$ \Lambda={\cal
O}_E\oplus u {\cal O}_E\oplus\cdots\oplus u^{n-1}{\cal O}_E,  $$ where
${\cal O}_E$ is the ring of integers of the field $E$. We note that $\OO_E$ is the unique
maximal order in $E$. In the so-called {\it Golden Division Algebra} (GDA)
\cite{BRV},
 i.e. the cyclic algebra $(E/F,\sigma,\gamma)$ obtained from the data
 $E={\Q}(i,\sqrt5)$, $F=\Q(i)$, $\gamma=i$, $n=2$, $\sigma(\sqrt5)=-\sqrt5$,
 the natural order $\Lambda$ is already maximal  \cite{HL}.  The ring of algebraic integers $\OO_E=\Z[i][\theta]$, when we denote the golden ratio by $\theta=\frac{1+\sqrt{5}}{2}$. The authors of \cite{BRV} further optimize the code by using an ideal $(\alpha)=(1+i-i\theta)$, and the Golden code is then defind as
 \begin{equation}
 \label{golden}
\mathcal{GC}=\left\{\frac{1}{\sqrt{5}}\left(
\begin{array}{cc}
\alpha x_0& i \sigma(\alpha)\sigma(x_1)\\
\alpha x_1&\sigma(\alpha)\sigma(x_0)\\
\end{array}\right)\ \Biggl|\ x_0,x_1\in\OO_E\right\}.
 \end{equation}
 The Golden code achieves the DMT as the element $\gamma=i$ is not in the image of the norm map. For the proof, see \cite{BRV}.
\end{exam}

\bigbreak
\begin{remark}
We feel that in \cite{BRV}, the usage of a maximal  order is just
a coincidence, as in this case it coincides with the natural order
which is generally used in ST code designs (cf. (\ref{cyclic
stbc})). At least the authors do not mention  maximal orders. As
far as we know, but our constructions (see also \cite{HLRV}) there does not exist any
designs using a maximal order other than the natural one.
 \end{remark}

\vspace*{4pt}

Next we  prove that the lattice
$L_6$ is optimal within the cyclic
division algebra $\HH$ in the sense that the diamond lattice $E_8=\hur(1+\xi)$  corresponds to a
proper ideal of a maximal order in $\HH$. \vspace*{4pt}

\begin{proposition}
\label{E8max} The ring $${\cal H}=\left\{q=c_1+\xi c_2+jc_3+j\xi c_4\in
\HH \ \vert\ c_1,\ldots,c_4\in \Q(i),\ (1+i)c_t\in
\G\ \forall t, c_1+c_3,c_2+c_4\in \G\right\}$$ is a maximal $\Z$-order
of the division algebra {\bf H}.
\end{proposition}

\vspace*{4pt}

\begin{proof}
Clearly the $\Q$-span of $\cal H$ is the whole algebra {\bf H},
and we have seen that ${\cal H}$ is a ring, so it is an order of
{\bf H}. Furthermore, if $\Lambda$ is any order of {\bf H}, then
so is $\Lambda[\sqrt2]=\Lambda\cdot {\bf \Z}[\sqrt2]$, as the
element $\sqrt2$ is in the center of {\bf H} (cf. Proposition
\ref{ronyaile 2.5}). Therefore it suffices to show that $\cal H$
is a maximal ${\bf \Z}[\sqrt2]$-order. In what follows, we will
call rational numbers in the coset $(1/2)+{\bf \Z}$ half-integers.
Assume for contradiction that we could extend the order $\cal H$ into a
larger order $\Gamma={\cal H}[q]$ by adjoining the quaternion
$q=a_1+a_2j$, where the coefficients
$$a_t=m_{t,0}+m_{t,1}\xi+m_{t,2}\xi^2+m_{t,3}\xi^3, \quad\hbox{\rm
$m_{t,\ell}\in{\Q}$ for all $t,\ell$}$$  are
elements of the field ${\Q}(\xi)$. As $\xi-\xi^3=\sqrt2$, and
$\xi^*=-\xi^3$, we see that  $$
tr(q)=a_1+a_1^*=2 m_{1,0}+\sqrt2 (m_{1,1}-m_{1,3}).
$$
 By Proposition \ref{normiehto} this must be an
element of ${\bf \Z}[\sqrt2]$, so we may conclude that $m_{1,0}$
must be an integer or a half-integer, and that $m_{1,1}-m_{1,3}$
must be an integer. Similarly  $$
tr(q\xi)=-2m_{1,3}+\sqrt{2}(m_{1,0}-m_{1,2}) $$
 must be an element
of ${\bf \Z}[\sqrt2]$. We may thus conclude that all the
coefficients $m_{1,\ell},\ \ell=0,1,2,3$ are integers or
half-integers, and that the pairs $m_{1,0}, m_{1,2}$ (resp.
$m_{1,1}, m_{1,3}$) must be of the same type, i.e. either both are
integers or both are half-integers. A similar study of $tr(q j)$
and $tr(qj\xi)$ shows that the same conclusions also hold for the
coefficients $m_{2,\ell},\ \ell=0,1,2,3$. Because ${\bf
\Z}[\xi]\subseteq{\cal H}$, replacing $q$ with any quaternion of
the form $q-\nu$, where $\nu\in \Z[\xi]$ will not change the
resulting order $\Gamma$. Thus we may assume that the coefficients
$m_{1,\ell},\ \ell=0,1,2,3$ all belong to the set $\{0,1/2\}$.
Similarly, if needed, replacing $q$ with $q-\nu' j$ for some
$\nu'\in\Z[\xi]$ allows us to assume that the coefficients
$m_{2,\ell},\ \ell=0,1,2,3$ also all belong to the set
$\{0,1/2\}$. Further replacements of $q$ by $q-\rho$ or
$q-\rho\xi$ then permit us to restrict ourselves to the case
$m_{2,\ell}=0$, for all $\ell=0,1,2,3$. If we are to get a proper
extension of $\cal H$, we are left with the cases $q=(1+i)/2$,
$q=\xi (1+i)/2$ and $q=(1+\xi)(1+i)/2$. We immediately see that
none of these have reduced norms in $\Z[\sqrt2]$, so we have
arrived at a contradiction.
\end{proof}

\begin{remark}
\label{icosian}
Another related well known maximal order is the icosian ring. It is a
maximal order in another subalgebra of the Hamiltonian quaternions, namely
$$ (\Q(i,\sqrt5)/\Q(\sqrt5),\sigma,-1),$$
where $\sigma$ is again the usual complex conjugation. This order
made a recent appearance as a building block of a MIMO-code in a
construction by Liu \& Calderbank. We refer the interested
reader to their work \cite{LC} or \cite{CS} for a detailed description of this order.
\end{remark}

The icosian ring and our order share one feature that is worth mentioning. As $2\times2$
matrices they do not have the non-vanishing determinant property. Algebraically this is a
consequence of the fact the respective centers, $\Q(\sqrt5)$ or $\Q(\sqrt2)$ both have
arbitrarily small algebraic integers, e.g. the sequence consisting of powers of the units
$(\sqrt5-1)/2$ (resp. $\sqrt2-1$) converges to zero. We shall return to this point in the next
section, where a remedy is described.

    \section{Decoding of the nested sequence of lattices}

\label{decoding} In this section, let us consider the coherent
MIMO case where the receiver perfectly knows the channel
coefficients. The received signal is $$\mathbf{y}=B
\mathbf{x}+\mathbf{n},$$ where $\bf{x}$ $\in\R^m$, $\bf{y},\ \bf{n}$
$\in\R^n$ denote the channel input, output and noise signals, and
$B\in\R^{n\times m}$ is the Rayleigh fading channel
response. The components of the noise vector $\bf{n}$ are i.i.d.
complex Gaussian random variables. In the special case of a MISO
channel ($n=1$),  the channel matrix takes a form of a vector
$B=\mathbf{h}\in\R^m$ (cf. Section \ref{intro}).

The information vectors  to be encoded into  our code matrices are
taken from the pulse amplitude modulation (PAM) signal set
$\mathcal{X}$ of the size $Q$, i.e., $$ \mathcal{X}=\{u=2q-Q+1\ \mid \
q\in\Z_Q\} $$ with $\Z_Q=\{0,1,...,Q-1\}$.

Under this assumption, the optimal  detector $g:\mathbf{y}\mapsto
\mathbf{\hat{x}}\in\mathcal{X}^m$ that minimizes the average error
probability $$ P(e)\stackrel{\Delta}{=}P(\mathbf{\hat{x}}\neq
\mathbf{x}) $$ is the maximum-likelihood (ML) detector given by
\begin{equation}
\label{detector}
\mathbf{\hat{x}}= \textrm{arg}\ \textrm{min}_{\mathbf{x}\in\Z_Q^m}\mid\mathbf{y-Bx}\mid ^2,
\end{equation}
where the components of the noise $\mathbf{n}$ have a common variance equal to one.

\subsection{Code controlled sphere decoding}
\label{CCSD}

The search in  (\ref{detector}) for the {\it closest lattice
point} to a given point $\mathbf{y}$ is known to be NP-hard in the
general case where the lattice does not exhibit any particular
structure. In \cite{P}, however, Pohst proposed an efficient
strategy of enumerating all the lattice points within a sphere
$\mathcal{S}(\mathbf{y},\sqrt{C_0})$ centered at $\mathbf{y}$ with a
certain radius $\sqrt{C_0}$ that works for lattices of a moderate
dimension. For background, see \cite{VB}-\cite{DGC}. For finite
PAM signals sphere decoders can also be visualized as a {\it
bounded search} in a tree.

The complexity of sphere decoders
critically depends on the preprocessing stage, the ordering in
which the components are considered, and the initial choice of the
sphere radius.  We shall use the standard preprocessing and
ordering that consists of the {\it Gram-Schmidt
orthonormalization} $B=(Q,Q')\begin{pmatrix}
  R \\
  0
\end{pmatrix}$ of the columns  of the channel matrix $B$
(equivalently, {\it $QR$ decomposition} on $B$) and the natural
back-substitution component ordering given by $x_m,...,x_1$. The
matrix $R$ is an $m\times m$ upper triangular matrix with positive
diagonal elements, $Q$ (resp. $Q'$) is an $n\times m$ (resp.
$n\times (n-m)$) unitary matrix, and $0$ is an $(n-m)\times m$ zero
matrix.

\def\midclose{\Bigr|}
\def\midopen{\Bigl|}
The condition  $B\mathbf{x}\in\mathcal{S}(\mathbf{y},$$\sqrt{C_0})$ can be written as
\begin{equation}
\label{in sphere}
\mid\mathbf{y}-B\mathbf{x}\mid ^2\leq C_0
\end{equation}
which after applying the $QR$ decomposition on $B$ takes the form
\begin{equation}
\label{in QR sphere}
\mid\mathbf{y'}-R\mathbf{x}\mid ^2\leq C_0',
\end{equation}
where $\mathbf{y'}=Q^T\mathbf{y}$ and
$C_0'=C_0-|(Q')^T\mathbf{y}|^2$.
Due to the upper triangular form of $R$, (\ref{in QR sphere}) implies the set of conditions
\begin{equation}
\label{QR conditions} \sum_{j=i}^m \midopen y_j'-\sum_{\ell=j}^m
r_{j,\ell} x_{\ell}\midclose ^2\leq C_0',\ \ \ i=1,...,m.
\end{equation}
The sphere decoding algorithm outputs the point $\mathbf{\hat{x}}$ for which the distance
\begin{equation}
\label{output}
d^2(\mathbf{y},B\mathbf{x})=\sum_{j=1}^m \midopen y_j'-\sum_{\ell=j}^m r_{j,\ell} x_{\ell}\midclose ^2
\end{equation}
is minimum. See details in \cite{DGC}.

The decoding of the base lattice $L_2$ can be performed  by using
the algorithm below proposed in \cite{DGC}. \vspace*{4pt}

\textbf{Algorithm II, Smart Implementation} (Input $C_0',\ \mathbf{y'},\ R$. Output $\mathbf{\hat{x}}.$)
\vspace*{4pt}

\textbf{STEP 1:} (Initialization) Set $i := m,\ T_m:=0,\ \xi_m:=0$, and $d_c:=C_0'$ (current sphere squared radius).
\vspace*{4pt}

\textbf{STEP 2:} (DFE on $x_i$) Set $x_i:=\left\lfloor (y_i'-\xi_i)/r_{i,i}\right\rceil$ and $\Delta_i:=sign(y_i'-\xi_i-r_{i,i}x_i)$.
\vspace*{4pt}

\textbf{STEP 3:} (Main step) If $d_c<T_i+\mid y_i'-\xi_i-r_{i,i}x_i\mid ^2$, then go to
STEP 4 (i.e., we are outside the sphere).

\hspace{1cm}Else if $x_i\notin\Z_Q$ go to STEP 6 (i.e., we are inside the sphere but outside the signal set boundaries).

\hspace{1cm}Else (i.e., we are inside the sphere and signal set boundaries) if $i>1$, then 

\hspace{1cm}\{let $\xi_{i-1}:=\sum_{j=i}^m r_{i-1,j}x_j,\ \ \  T_{i-1}:=T_i+\mid y_i'-\xi_i-r_{i,i}x_i\mid ^2,\ \ \ i:=i-1$, and go to STEP 2\}.

\hspace{1cm}Else (i=1) go to STEP 5.
\vspace*{4pt}

\textbf{STEP 4:} If $i=m$, terminate, else set $i:=i+1$ and go to STEP 6.
\vspace*{4pt}

\textbf{STEP 5:} (A valid point is found) Let $d_c:=T_1+\mid y_1'-\xi_1-r_{1,1}x_1\mid ^2$, save $\mathbf{\hat{x}}:=\mathbf{x}$. 

\hspace{1cm}Then, let $i:=i+1$ and go to STEP 6.
\vspace*{4pt}

\textbf{STEP 6:} (Schnorr-Euchner enumeration  of level $i$) Let $x_i:=x_i+\Delta_i,\ \Delta_i:=-\Delta_i-sign(\Delta_i)$.

\hspace{1cm}Then, go to STEP 3.
\vspace*{4pt}

Note that  given the values $x_{i+1},...,x_m$, taking the ZF-DFE (zero-forcing decision-feedback equalization) on  $x_i$ avoids retesting other nodes at level $i$ in case we fall outside the sphere.  Setting $d_c=\infty$ would ensure that the first point found by the algorithm is the ZF-DFE  point (or the Babai point) \cite{DGC}. However, if the distance between the ZF-DFE point and the received signal is very large this choice may cause some inefficiency, especially for high dimensional lattices.

The decoding of  the other three lattices in (\ref{nested
sequence}) also relies  on this algorithm, but we need to run some
additional parity checks. This  simply means that in addition to
the checks concerning the facts that we have to be both inside the
sphere radius and inside the signal set boundaries, we also have
to lie inside a given sublattice. This will be taken care of by
a method we call {\it code controlled sphere decoding} (CCSD), that combines the
algorithm above with certain case considerations. To this end, let
us write the constraints on the elements $c_i$ as  {\it modulo $2$
operations}. Denote by $\mathbf{x}=(x_1,x_2,...,x_8)=(\Re c_1, \Im
c_1,...,\Re c_4, \Im c_4)\in\R^8$ the real vector corresponding to
the channel input.  Note that when exploiting these
relations in the CCSD algorithm, we have to use different
orderings for the basis matrices of the lattice in different cases
in order to make the parity checks as simple as possible. Let us
first order the basis matrices as $B_1=M(1,0,0,0),
B_2=M(i,0,0,0),...,B_7=M(0,0,0,1),B_8=M(0,0,0,i)$. Then when
decoding e.g. the $L_5$ lattice, we reorder the basis matrices as
$B_1,B_2,B_5,B_6,B_3,B_4,B_7,B_8$ in order to get the sum
$c_1+c_3$ as the sum of the first $4$ components and the sum
$c_2+c_4$ as the sum of the last $4$ components (cf. Proposition
\ref{D4D4}). The conditions for the Gaussian elements of
Propositions \ref{D8}-\ref{E8} can clearly be translated into the
following  modulo $2$ integer conditions, see for instance  Remark
\ref{D8 remark}. The additional parity check steps will hence be
as shown  in Table \ref{cases} above.

\def\korkeutta{\hbox{\vrule height1.1\baselineskip
    depth.3\baselineskip
    width0pt}}
\begin{table}
\caption{CCSD: Additional case considerations}
\label{cases}
\begin{center}
\begin{tabular}{|l|l|}
\hline
\large \korkeutta CASE $L_4$ &  \large $\sum_{i=1}^8 x_i\equiv 0\
(\textrm{mod}\ 2)$\\[1.5mm] \hline
\large \korkeutta CASE $L_5$ & \large
$x_1+x_2\equiv x_5+x_6$,\\ & \large $x_3+x_4\equiv x_7+x_8\ (\textrm{mod}\ 2)$\\[1.5mm] \hline
\large \korkeutta CASE $L_6$ &  \large $x_1+x_2\equiv x_3+x_4 \equiv
x_5+x_6 \equiv x_7+x_8,$\\
& \large $\sum_{2\mid i} x_i \equiv \sum_{2\nmid i} x_i\equiv 0\ (\textrm{mod}\ 2)$\\[1.5mm]
\hline
\end{tabular}
\end{center}
\end{table}

\begin{figure*}[!ht]
\begin{center}
\includegraphics[height=7.3cm]{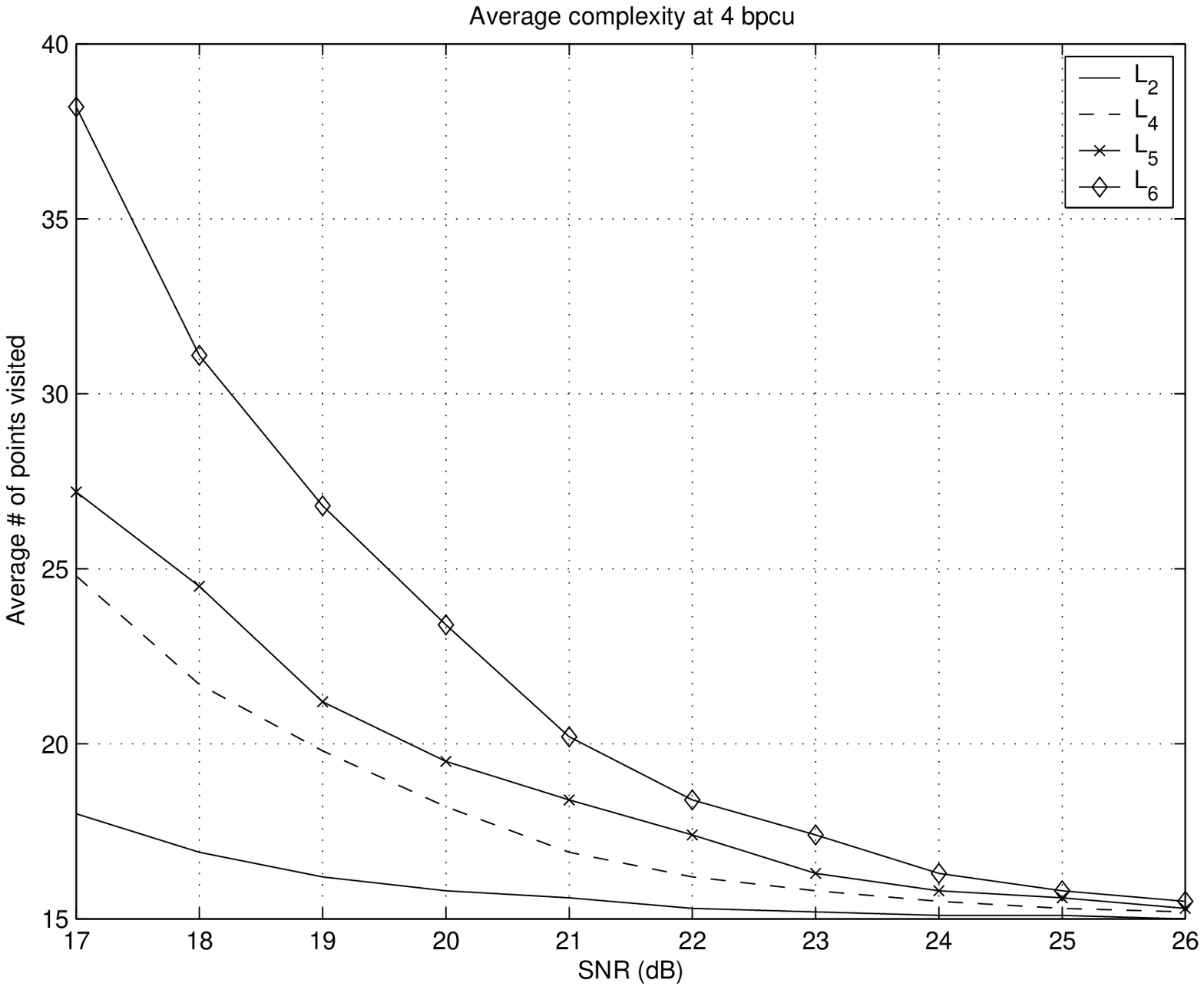}
\includegraphics[height=7.3cm]{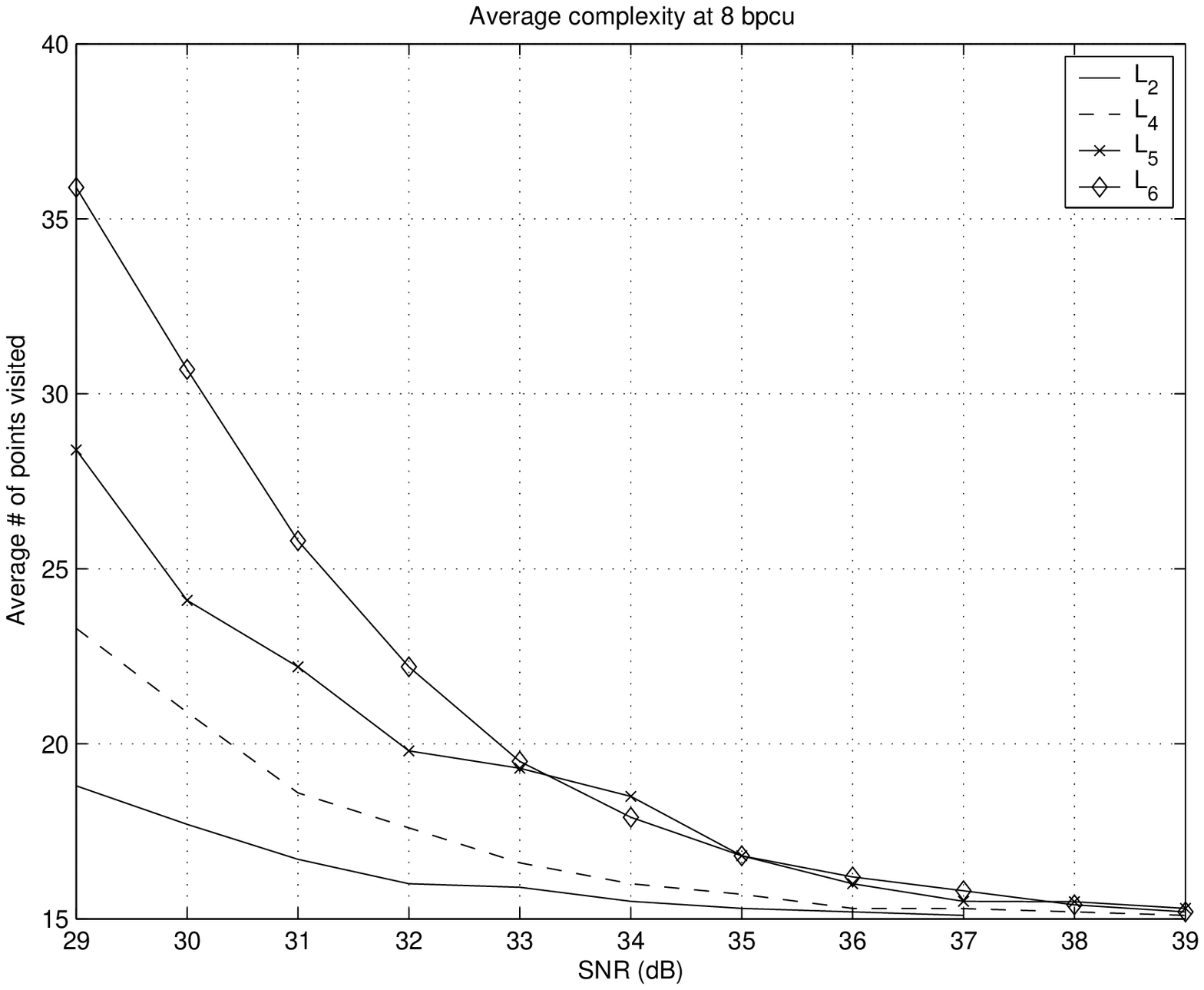}
\caption{Average complexity of $4$ tx-antenna  matrix lattices at
rates (approximately) $R=4$ and $R=8$ bpcu.}\label{fig1}
\end{center}
\end{figure*}


 As the Alamouti scheme \cite{Alam} has a very
efficient decoding algorithm available, and our quaternionic
lattices have an Alamouti-like block structure, it is natural to
ask whether  any of the benefits of Alamouti decoding will survive
for our lattices. We shall see that the block structure allows us
to decode the two blocks independently from each other. The
following simple observation is the underlying geometric reason
for our ability to do this. \vspace*{4pt}
\begin{lemma}
\label{lohkoortogonaalisuus}
Let $A$ and $B$ be two $n\times n$ matrices with the property that
the matrices $A,B,A^H,B^H$ commute. Let $\mathbf{h}\in \C^{2n}$ be any (row) vector
and write
$$
M(A,B)=\left(
\begin{array}{cc}
A&B\\ -B^H&A^H\\
\end{array}\right).
$$ Then the vectors $\mathbf{h} M(A,0)$ and $\mathbf{h} M(0,B)$ are orthogonal to
each other when we identify $\C^{2n}$ with $\R^{4n}$ and use the
usual inner product of a vector space over the real numbers.
\end{lemma}
\vspace*{4pt}
\begin{proof}
With the identification $\C^{2n}=\R^{4n}$ the real inner product
is the real part of the hermitian inner product $\langle\ ,\
\rangle$ of $\C^{2n}$. Write the vector $\mathbf{h}$ in the block form
$\mathbf{h}=(h^{(1)}, h^{(2)})$, where the blocks $h^{(j)}, j=1,2$, are
(row) vectors in $\C^n$. Then we can compute
\begin{eqnarray*}
&&\langle \mathbf{h} M(A,0), \mathbf{h} M(0, B)\rangle\\
&=&\langle \mathbf{h} M(A,0) M(0,B)^H,\mathbf{h}\rangle\\
&=&\langle \mathbf{h} M(A,0) M(0,-B), \mathbf{h}\rangle\\
&=&\langle \mathbf{h} M(0,-AB),\mathbf{h}\rangle\\
&=&\langle h^{(2)} A^HB^H, h^{(1)}\rangle
-\langle h^{(1)} AB, h^{(2)}\rangle.
\end{eqnarray*}
As $\langle \mathbf{u} M, \mathbf{v}\rangle= \langle \mathbf{v} M^H, \mathbf{u}\rangle^*$
for all vectors $\mathbf{u},\mathbf{v}$ and matrices $M$, we see that the
above hermitian inner product is pure imaginary.
\end{proof}
\vspace*{4pt}
\begin{corollary}
Let $A$ and $B$ range over sets of $n\times n$-matrices. Let $\mathbf{h}$
and $\mathbf{r}$ be vectors in $\C^{2n}$. Then the Euclidean distance
between $\mathbf{r}$ and $\mathbf{h} M(A,B)$ is minimized for the
 $A=A_0$
and $B=B_0$, when $A_0$ minimizes the Euclidean distance between
$\mathbf{r}$ and $\mathbf{h} M(A,0)$ and $B_0$ minimizes the Euclidean distance
between $\mathbf{r}$ and $\mathbf{h} M(0,B)$.
\end{corollary}
\vspace*{4pt}
\begin{proof}
Write $V_A$ (resp. $V_B$) for the real vector space spanned by the
vectors $\mathbf{h} M(A,0)$ (resp.  $\mathbf{h} M(0,B)$). These subspaces are
orthogonal to each other in the sense of Lemma
\ref{lohkoortogonaalisuus}. Whence we can uniquely write
$\mathbf{r}=r_A+r_B+r_{\perp}$, where $r_A\in V_A, r_B\in V_B$ and
$r_{\perp}$ is in the (real) orthogonal complement of the direct
sum $V_A\oplus V_B$. A similar decomposition for the vector $\mathbf{h}
M(A,B)$ is $\mathbf{h} M(A,B)=h_A+h_B$, where $h_A=\mathbf{h} M(A,0)\in V_A$ and
$h_B=\mathbf{h} M(0,B)\in V_B$. By the Pythagorean theorem $$ \abs{\mathbf{r}-\mathbf{h}
M(A,B)}^2=\abs{r_A- \mathbf{h} M(A,0)}^2+ \abs{r_B- \mathbf{h}
M(0,B)}^2+\abs{r_{\perp}}^2. $$ Furthermore, here $$\abs{r_A- \mathbf{h}
M(A,0)}^2=\abs{\mathbf{r}- \mathbf{h} M(A,0)}^2-\abs{r_B}^2-\abs{r_{\perp}}^2,$$ so
the quantities $\abs{r_A- \mathbf{h} M(A,0)}^2$ and $\abs{\mathbf{r}- \mathbf{h} M(A,0)}^2$
are minimized for the same choice of the matrix $A$. A similar
argument applies to the $B$-components, so the claim follows.
\end{proof}

\subsection{Complexity issues and collapsing lattices}
\label{complexity}

The number of nodes in the search tree is used as a measure of complexity so that the
implementation details or the physical environment do not affect it.
We have analyzed many different kinds of situations concerning  the change of complexity
of the sphere decoder when moving in (\ref{nested sequence}) from  right to left.

In Fig. \ref{fig1} we have plotted the average number of points visited by the algorithm
in different cases at the rates approximately $4$ and $8$ bpcu. The SNR regions cover the
block error rates between  $\approx 10\%-0.01\%$. As can be seen, in the low SNR end, the
difference in complexity between the different lattices is  clear but evens out when the SNR
increases. For the sublattices $L_4,\ L_5$, and $L_6$ the algorithm  visits  $1.1-2.1$ times
as many points as for the base lattice $L_2$.  In the larger SNR end, the performance is
fairly similar for all the lattices. E.g. at $4$ and $8$ bpcu, when all the lattices reach
the bound of maximum 20 points visited, the block error rates of $L_4,\ L_5$, and $L_6$ are
still as big as $5\%,\ 2\%$, and $1\%$ respectively.

\vspace*{4pt}
\begin{definition}
\label{defect}
In a MISO setting we say that a matrix lattice $L$ of rank $m$ {\em collapses at a channel realization $\mathbf{h}$},
if the receiver's version of the lattice $\mathbf{h}L$ spans a real vector space of dimension $<m$. We call the set of
such channel realizations the critical set. We say that the {\it sensitivity} $s(L)$
(towards collapsing) of the lattice $L$ is $r$, if the critical set is a union of finitely many
subspaces of real dimension $\le r$.
\end{definition}
\vspace*{4pt}

So we e.g. immediately see that a lattice residing in an orthogonal design will have zero sensitivity.
While we have no precise results the thinking underlying the concept can be motivated as follows. When the
infinite lattice collapses into a lower dimensional space, its linear structure is severely mutilated. For
example the minimum Euclidean distance drops to zero --- for any $\epsilon>0$ there will be infinitely many
other lattice points within a distance $<\epsilon$. Even when we restrict ourselves to a finite subset of
the lattice, the coordinates of the nearby points may differ drastically. Thus even an ML-decoder will have
problems, and an algorithm relying on the orderly linear structure of the lattice (like the sphere decoder)
cannot work very efficiently. Similar problems are still there, when the actual channel realization $\mathbf{h}$
is close to a critical vector.

The sensitivity then enters the scene as a crude measure for the probability of this happening. It is easy to
see that in a Rayleigh fading channel the probability of the channel vector $\mathbf{h}$ to be within $\epsilon$
of a critical vector behaves like ${\cal O}(\epsilon^{2n-s})$. Thus the lower the sensitivity, the lower
the probability of the lattice becoming distorted by the channel.

We lead off by determining the sensitivity of the DAST-lattices.

\vspace*{4pt}
\begin{exam}
\label{dast}
There exist 8-dimensional lattices \cite{DAB} of $4\times4$ matrices of the form
$$M_{DAST}=
\left(
\begin{array}{rrrr}
x_1&x_2&x_3&x_4\\
x_1&-x_2&x_3&-x_4\\
x_1&x_2&-x_3&-x_4\\
x_1&-x_2&-x_3&x_4\\
\end{array}\right).$$
These matrices are simultaneously diagonalizable as they have common orthogonal eigenvectors
$\mathbf{h}_1=(1,1,1,1)$, $\mathbf{h}_2=(1,-1,1,-1)$, $\mathbf{h}_3=(1,1,-1,-1)$
and $\mathbf{h}_4=(1,-1,-1,1)_4$. Write the channel vector in terms of this
basis $\mathbf{h}=\sum_{j=1}^4a_j\mathbf{h}_j$. If any of the coefficients vanishes, say $a_k=0$, then
the DAST-lattice collapses, because the receiver's version of the lattice will belong to
the complex span of the other three eigenvectors $\mathbf{h}_j, j\neq k$. On the other hand, if all
the coefficients $a_j\neq0, j=1,2,3,4$, this channel vector will not be critical. One way of seeing this is that applying the linear mapping
determined by $\mathbf{h}_j\mapsto (1/a_j)\mathbf{h}_j$ to the receiver's lattice then recovers the
original full rank lattice of vectors $(x_1,x_2,x_3,x_4)$. Such a mapping obviously cannot affect
the dimension of the space spanned by the vectors, so the lattice won't collapse.

We have shown that the sensitivity of the DAST-lattice is six.
\end{exam}
\vspace*{4pt}

\begin{figure*}[!ht]
\begin{center}
\includegraphics[height=7.3cm]{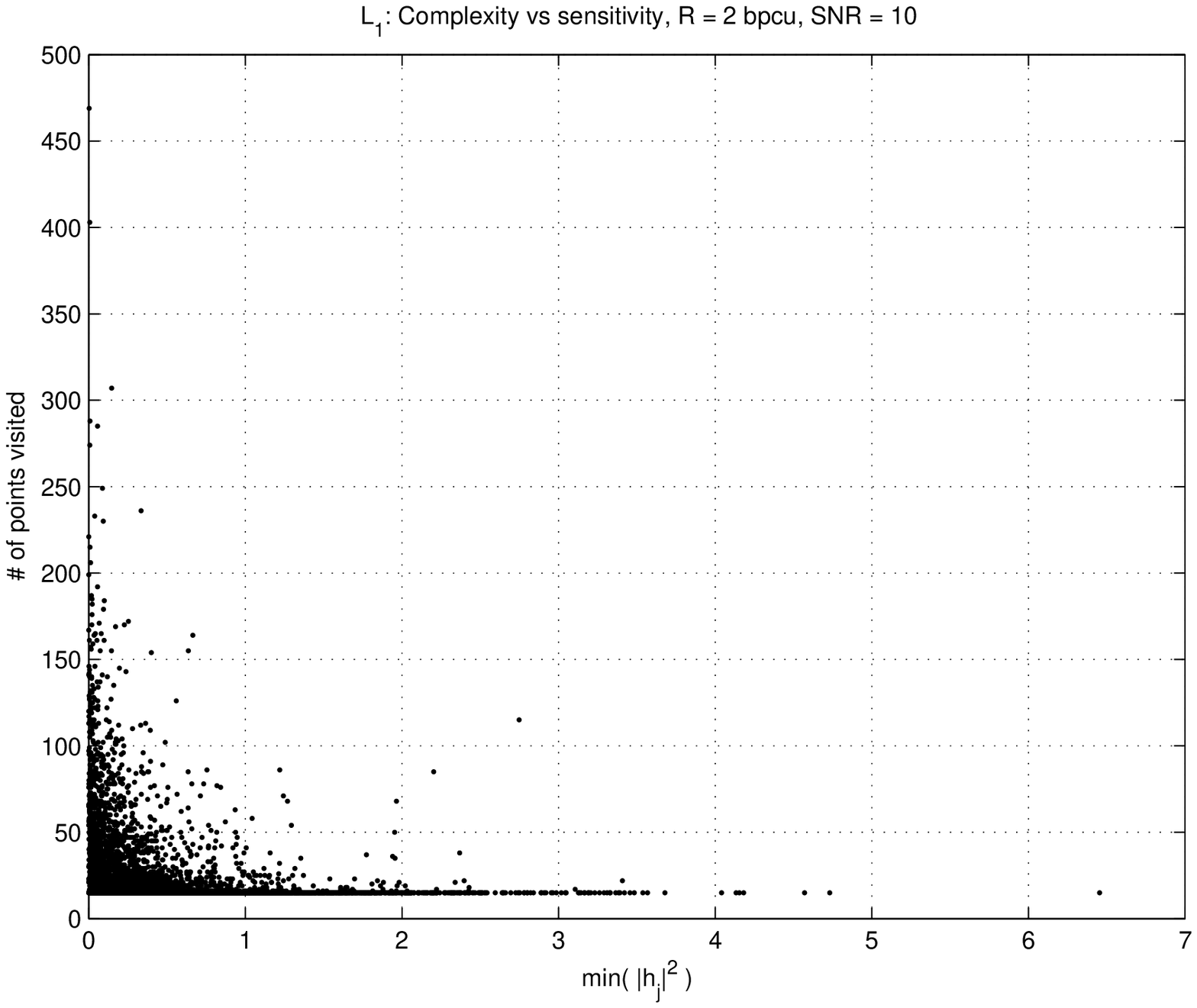}
\includegraphics[height=7.3cm]{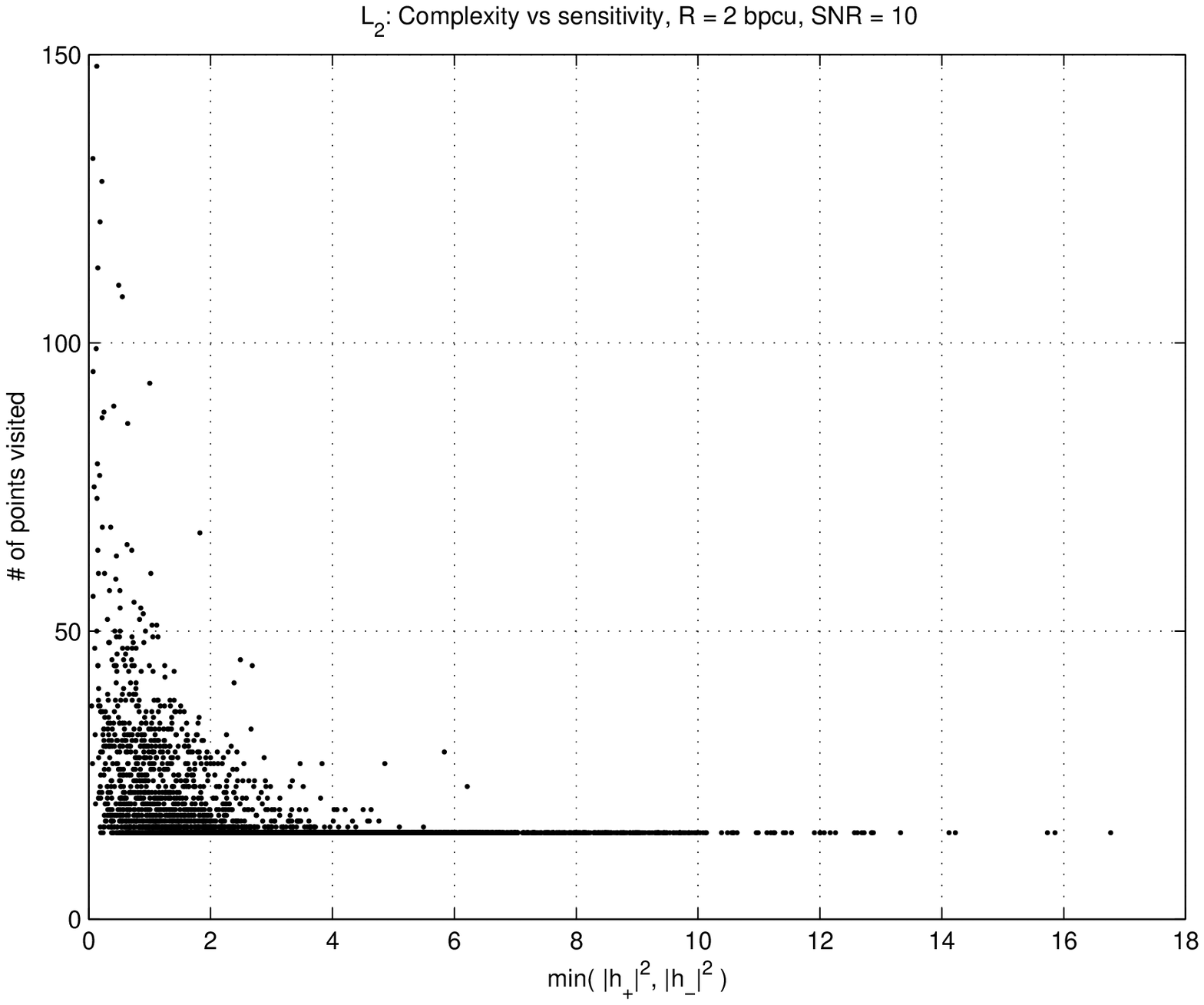}
\includegraphics[height=7.3cm]{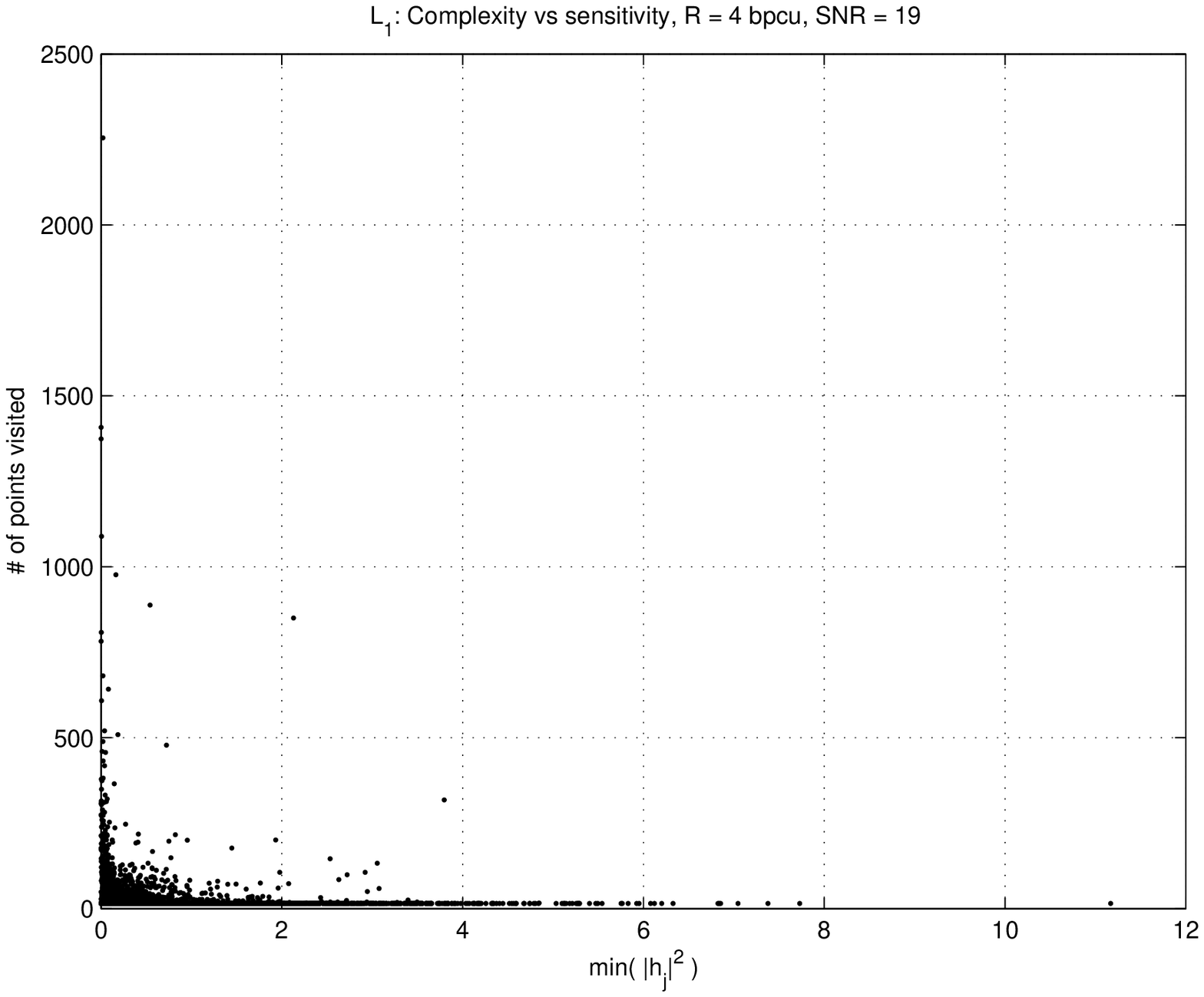}
\includegraphics[height=7.3cm]{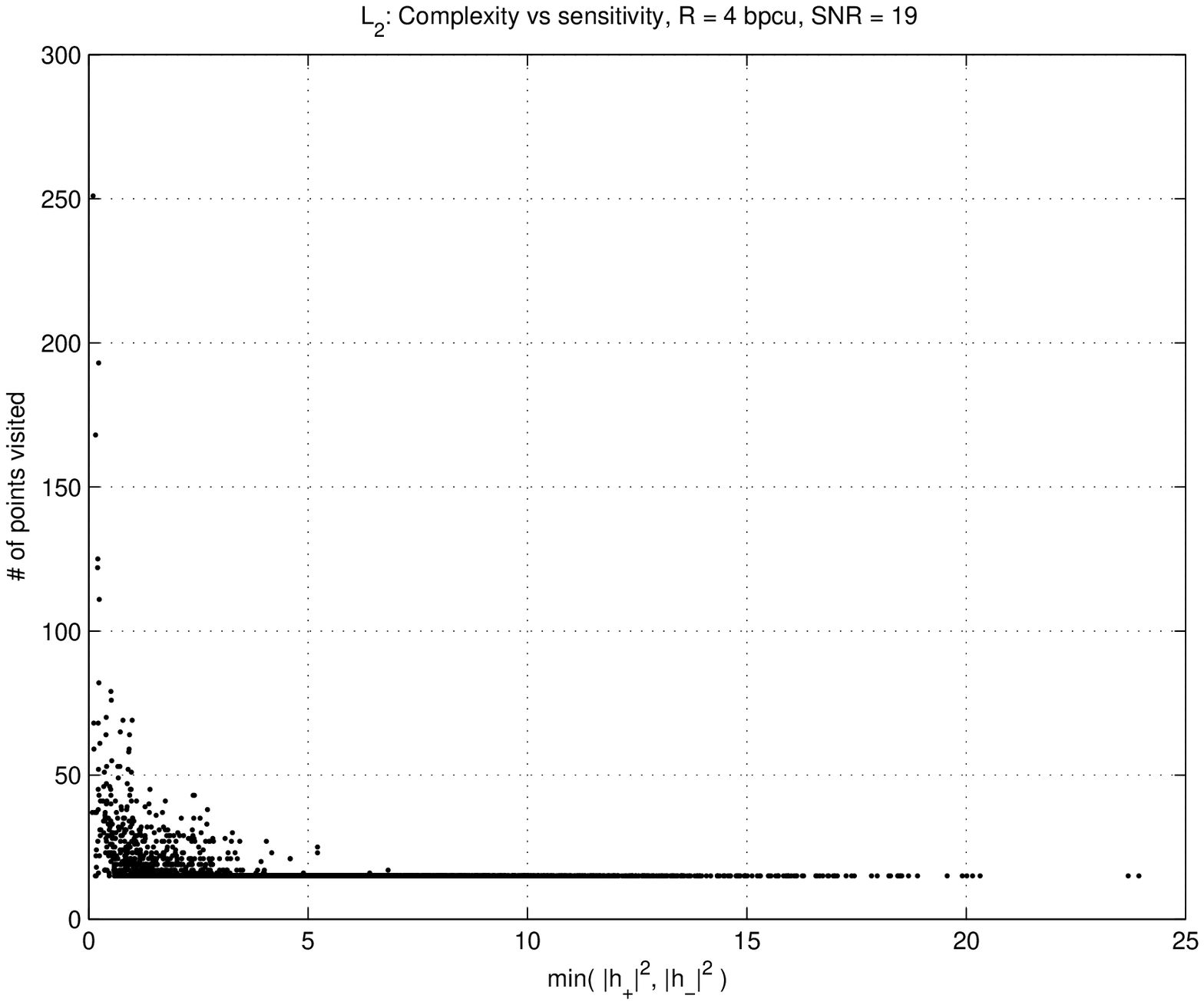}
\caption{The impact of sensitivity on complexity, $L_1\ (\approx L_{DAST})$ vs $L_2$.}\label{fig2}
\end{center}
\end{figure*}

We proceed to determine the sensitivities
of the lattices $L_1$ of Proposition \ref{rings} and the ones within the nested sequence (\ref{nested sequence}). Let us first consider $L_1$.
Let $$U=\left(
\begin{array}{rrr}
\mathbf{h}_1\\
\vdots\\
\mathbf{h}_4\\
\end{array}\right)$$ be the $4\times4$ matrix
with rows $\mathbf{h}_1,\mathbf{h}_2,\mathbf{h}_3,\mathbf{h}_4$ of the form $(1,\zeta^j,\zeta^{2j},\zeta^{3j})$ for
$j=1,5,9,13$. Recall that earlier we have used $\{1,\zeta,\zeta^2,\zeta^3\}$
as an integral basis, so the rows of $U$ are the images of this ordered
basis under the action of the Galois group $G$ of the extension
$\Q(\zeta)/\Q(i)$. Now it happens that the matrix $U$ is unitary
(up to a constant factor) as $UU^*=4I_4$.
Let $z=c_1+c_2\zeta+c_3\zeta^2+c_4\zeta^3$ be an arbitrary algebraic
integer of $\Q(\zeta)$, and $M(z)=M_L(c_1,c_2,c_3,c_4)\in L_1$ be the
corresponding matrix of Proposition \ref{rings}.
 According to the
theory of algebraic numbers (and also trivially verified by hand)
the rows of $U$ are (left) eigenvectors of $M(z)$, and
$$
U M(z) U^{-1}=
\left(
\begin{array}{cccc}
z&0&0&0\\
0&\sigma_2(z)&0&0\\
0&0&\sigma_3(z)&0\\
0&0&0&\sigma_4(z)
\end{array}\right)
$$
is a diagonal matrix with entries gotten by applying the elements of
the Galois group $G=\{\sigma_1=id,\sigma_2,\sigma_3,\sigma_4\}$
to the number $z$.

So all the matrices $M_L(c_1,c_2,c_3,c_4)$ are diagonalized by $U$.
Therefore we might call the lattice $L_1$ `DAST-like', as it shares this
property with the lattices from \cite{DAB}.
\vspace*{4pt}

\begin{proposition}
\label{nfdefect}
The lattice $L_1$ has sensitivity six.
\end{proposition}
\vspace*{4pt}

\begin{proof}
The situation is completely analogous to that of Example \ref{dast}.
The lattice $L_1$ will collapse, iff the channel realization belongs to
any of the 4 complex vector spaces spanned by any three of the common
eigenvectors.
\end{proof}
\vspace*{4pt}

\begin{figure*}[!ht]
\begin{center}
\includegraphics[height=7.3cm]{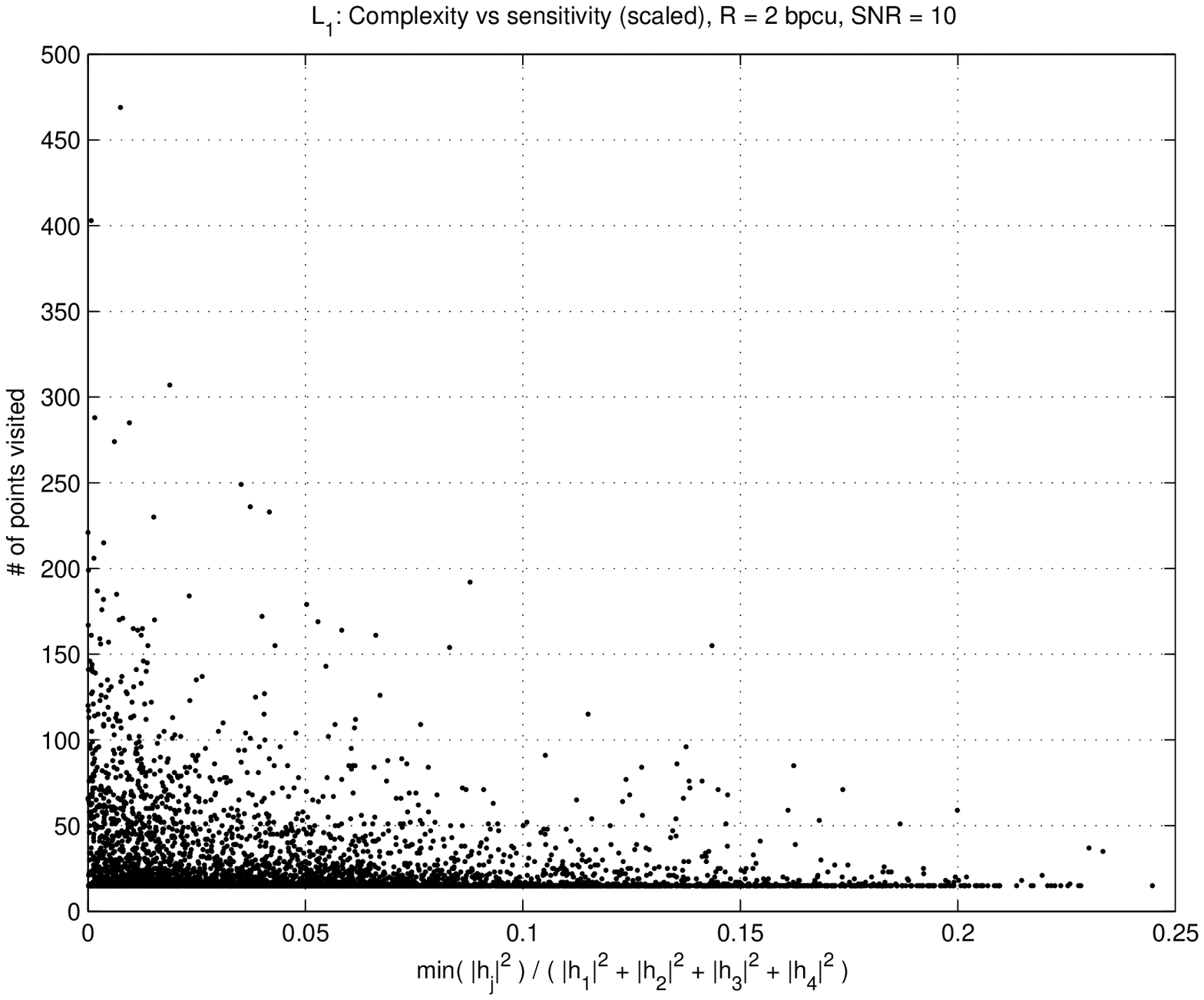}
\includegraphics[height=7.3cm]{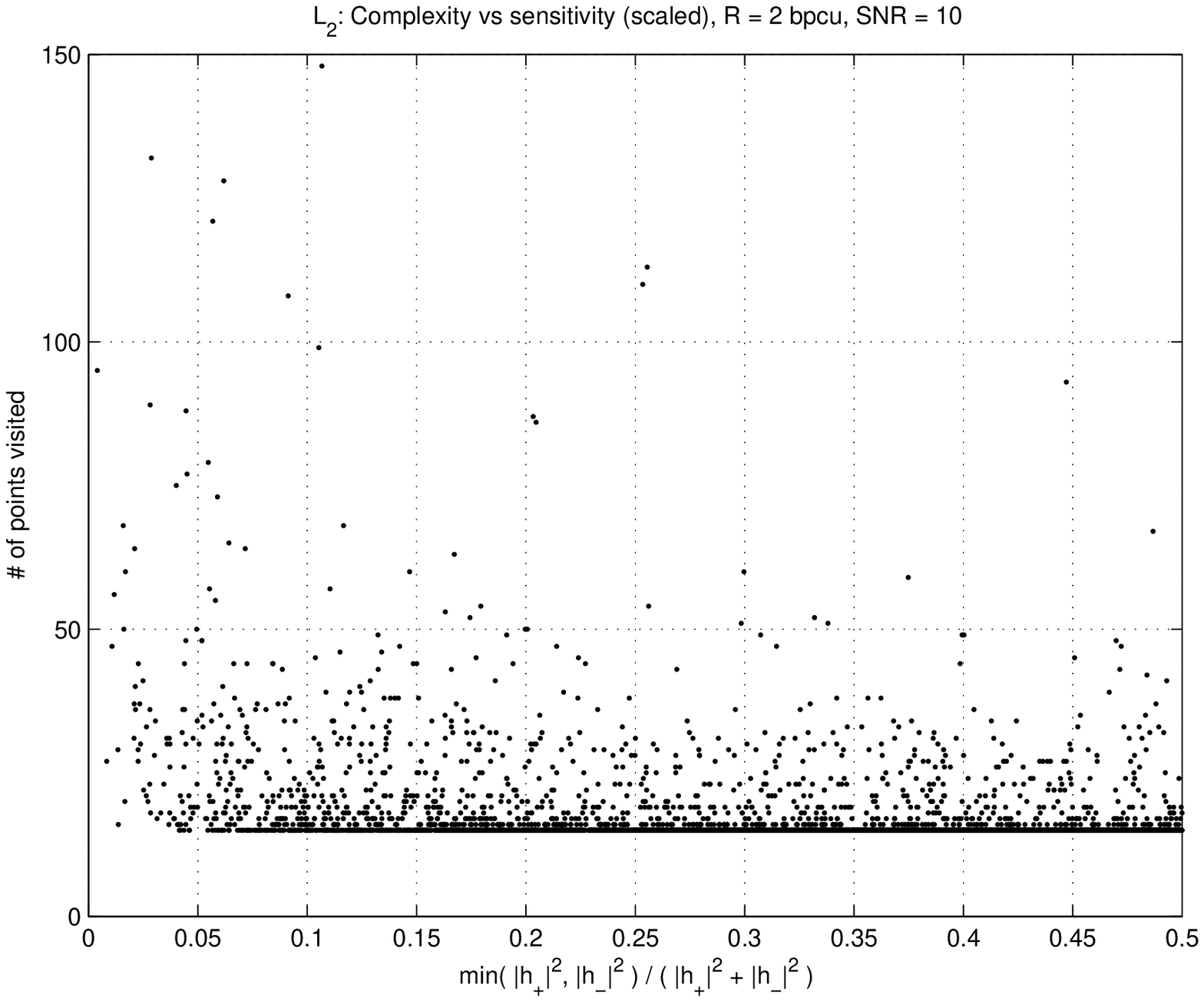}
\includegraphics[height=7.3cm]{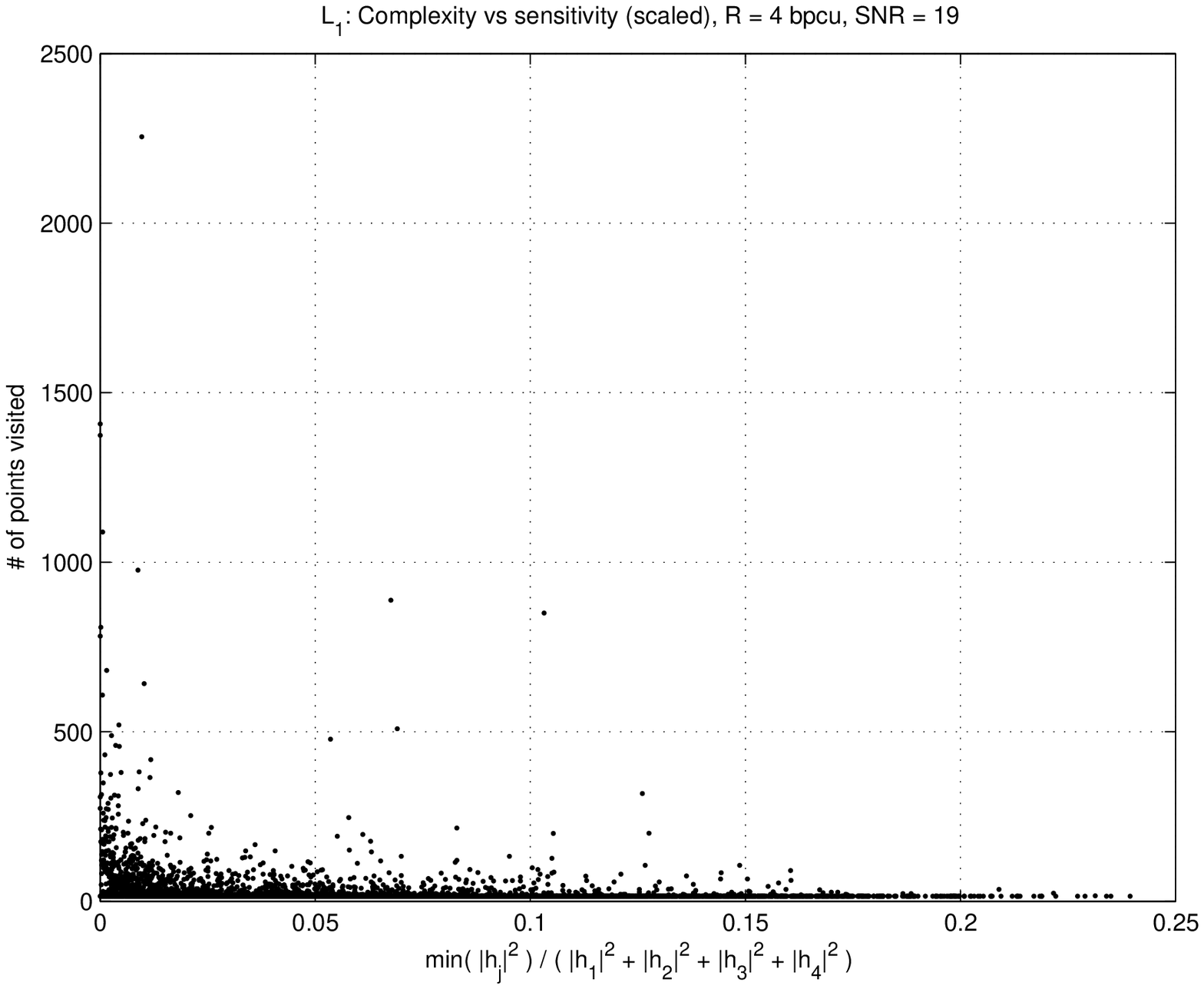}
\includegraphics[height=7.3cm]{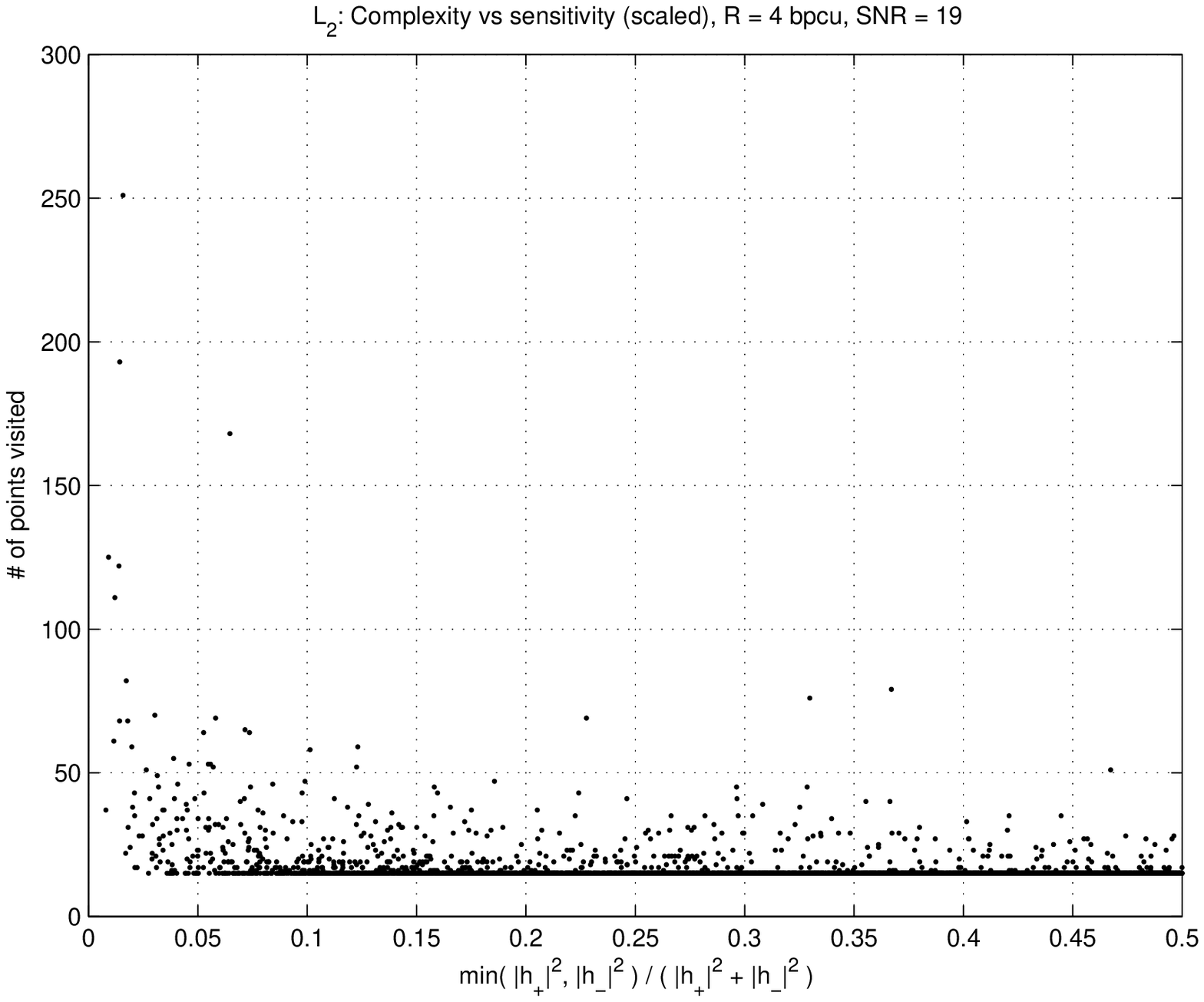}
\caption{The scaled impact of sensitivity on complexity, $L_1\ (\approx L_{DAST})$ vs $L_2$.}\label{fig3}
\end{center}
\end{figure*}

In order to study the quaternionic lattices we first
observe that the $2\times2$-matrices $A$ and $B$ appearing as
blocks of a matrix $M\in L_2$
all have $(1,\pm\xi)$ as their common (left) eigenvectors.
The same holds for the adjoints $A^*, B^*$ as they also appear
as blocks of $M^*$ that also happens to belong to the lattice $L_2$.
From the proof of Proposition \ref{D8} we see that
the matrix $MM^*$, $M=M(c_1,c_2,c_3,c_4)$, has eigenvalues
$\alpha\pm\abs{k}$ with respective (left) eigenvectors
$(1,\pm\xi,0,0)$ and $(0,0,1,\pm\xi)$. Here
$\alpha=\sum_{j=1}^4\abs{c_j}^2$
and $k=-ic_1c_2^*+c_2c_1^*-ic_3c_4^*+c_4c_3^*$. We make this
more precise before we determine the sensitivity of the quaternionic lattices.

There is a connection between our MISO-code and the multi-block codes introduced
by Belfiore in \cite{YB} and Lu in \cite{L} that can be best explained
with the notation of the present section.
Consider the unitary matrix with the above basis vectors as columns
$$
U=\frac{1}{\sqrt2}\left(
\begin{array}{cccc}
1 & 1 & 0 & 0 \\
\xi & -\xi & 0 & 0 \\
0 & 0 & 1 & 1 \\
0 & 0 & \xi & -\xi
\end{array}\right).
$$
If we conjugate the matrices of the algebra $\HH$ by $U$ we get matrices of the form
$$
\left(
\begin{array}{cccc}
x_1 & -x_2^* & 0 & 0 \\
x_2 & x_1^* & 0 & 0 \\
0 & 0 & \tau(x_1) & -\tau(x_2) \\
0 & 0 & \tau(x_2) & \tau(x_1)^*
\end{array}\right),
$$
where the elements $x_1, x_2$ belong to the field $\Q(\xi)=\Q(i,\sqrt2)$, and
$\tau:\Q(\xi)\rightarrow\Q(\xi)$ is the automorphism determined by $\tau(i)=i$,
$\tau(\sqrt2)=-\sqrt2$. Thus we see that our MISO-code is unitarily equivalent
to a multi-block code with a structure similar to \cite{L} --- only our center
is smaller.

The upshot here, as well as in \cite{YB}, \cite{L}, and in the icosian
construction from \cite{LC} is that while the individual diagonal blocks may have arbitrarily
small determinants, when we use them together with their algebraic conjugates, the diagonal blocks
together conspire to give a non-vanishing determinant. This is because the algebraic conjugates of
small numbers are necessarily just large enough to compensate as the algebraic norms are known to be
integers.

Another benefit enjoyed by our matrix representation of the algebra $\HH$ over the above
multi-block representation is that the signal constellation is better behaved. Surely the simple
QAM-constellation of our matrices is to be preferred over the linear combinations of two rotated
QAM-symbols of the multi-block representation.

This feature clearly begs to be generalized to a MIMO-setting. One such construction is the
previously mentioned icosian construction of Liu \& Calderbank \cite{LC}, where they
managed to add a multiplexing gain of 2 to a similar multi-block representation of the icosians.
It turned out that the question of how to best do this in the spirit of the present article is
somewhat delicate. The resulting codes will necessarily be asymmetric MIMO-codes, and we refer
the reader to  \cite{HLu}.

We return to the sensitivity of the quaternionic lattices. The following result is now
easy to verify

\begin{proposition}
\label{qnfdefect}
Let $V_+$ (resp. $V_-$) be the complex subspace of $\C^4$ generated
by the vectors $(1,\xi,0,0)$ and $(0,0,1,\xi)$ (resp. by $(1,-\xi,0,0)$ and $(0,0,1,-\xi)$). The subspaces $V_+$ and $V_-$ are orthogonal
complements of each other in $\C^4$, so any channel vector
can be uniquely written as
 $$\mathbf{h}=\mathbf{h}_++\mathbf{h}_-,$$ where $\mathbf{h}_{\pm}\in V_{\pm}$ respectively. If $\mathbf{h}$ belongs
to one of the subspaces $V_+, V_-$, the lattice $\mathbf{h}L_2$ collapses. Otherwise the lattice $L_2$
does not collapse. In particular the sensitivity of the lattices $L_2, L_3, L_4, L_5, L_6$ is four.
\hfill  \QED
\end{proposition}

\vspace*{4pt}

Our simulations, indeed, show that
the complexity of a sphere decoder increases sharply, when we
approach the critical set.  A comparison between the lattices $L_1$
and $L_2$ does not show a dramatic difference between the average
complexities of a sphere decoder, but the difference becomes very
apparent, when studying the high-complexity tails of the complexity
distribution.

In Fig. \ref{fig2} we have plotted the complexity distribution of 5000 transmissions
for different data rates. On the horizontal axis the quantity min$(\ |\mathbf{h}_i|^2\ )$
(resp. min$(\ |\mathbf{h}_+|^2,|\mathbf{h}_-|^2\ )$) describes how close the
lattice $L_1$ (resp. $L_2$) is to the situation where it would collapse. That
is, how close to zero the minimum of the components $\mathbf{h}_i\in V_i,\  i=1,2,3,4$,
(resp. $\mathbf{h}_{\pm}\in V_{\pm}$) gets (cf. Remark \ref{nfdefect} and
Proposition \ref{qnfdefect}). For both $L_1$ and $L_2$ the figure shows that
the smaller the quantity, the higher the complexity. We can also conclude that the
lattice $L_1$  nearly collapses a lot more often than the lattice $L_2$. In addition, the
number of points visited by the sphere decoding algorithm is much higher for $L_1$ than for $L_2$.
These are phenomena caused by the higher sensitivity of $L_1$. In Fig. \ref{fig3} the scaled
impact of sensitivity is depicted.

Note that as $L_{DAST}$ has the same sensitivity as $L_1$, we can equally well analyze the behavior of the DAST lattice
on the basis of Fig. \ref{fig2} and Fig. \ref{fig3}.

\section{Energy considerations and simulations}
\label{simulation}
As a summary of Propositions \ref{rings}--\ref{E8} we get the following.
\vspace*{4pt}
\begin{proposition}
\label{pdmin}
(1) The lattice $L_2$ is isometric to the rectangular lattice $\Z^8$
and has a minimum determinant equal to $1$.

(2) The lattice $L_4$ isometric to $D_8$ is an index two sublattice of $L_2$ and has a
minimum determinant equal to $4$.

(3) The lattice $L_5$ isometric to $D_4\bot D_4$ is an index four sublattice of $L_2$ and
has a minimum determinant equal to $16$.

(4) The lattice $L_6$ isometric to $E_8$ is an index 16 sublattice of $L_2$ and has a
minimum determinant equal to $64$.\hfill  \QED
\end{proposition}
\vspace*{4pt}

 In order to compare these lattices we scale them
to the same minimum determinant. When a real scaling factor
$\rho$ is used the minimum determinant is multiplied by
$\rho^2$. As all the lattices have rank 8, the fundamental volume
is then multiplied by $\rho^8$. Let us choose the units so that
the fundamental volume of $L_2$ is $m(L_2)=1$. Then after scaling
$m(L_4)=1/2$, $m(L_5)=1/4$, and $m(L_6)=1/4$. As the density of a
lattice is inversely proportional to the fundamental volume, we
thus expect the codes constructed within e.g. the lattices $L_4$
and $L_6$ to outperform the codes of the same size within $L_2$.

\begin{figure}[!ht]
\begin{center}
\includegraphics[height=7.2cm]{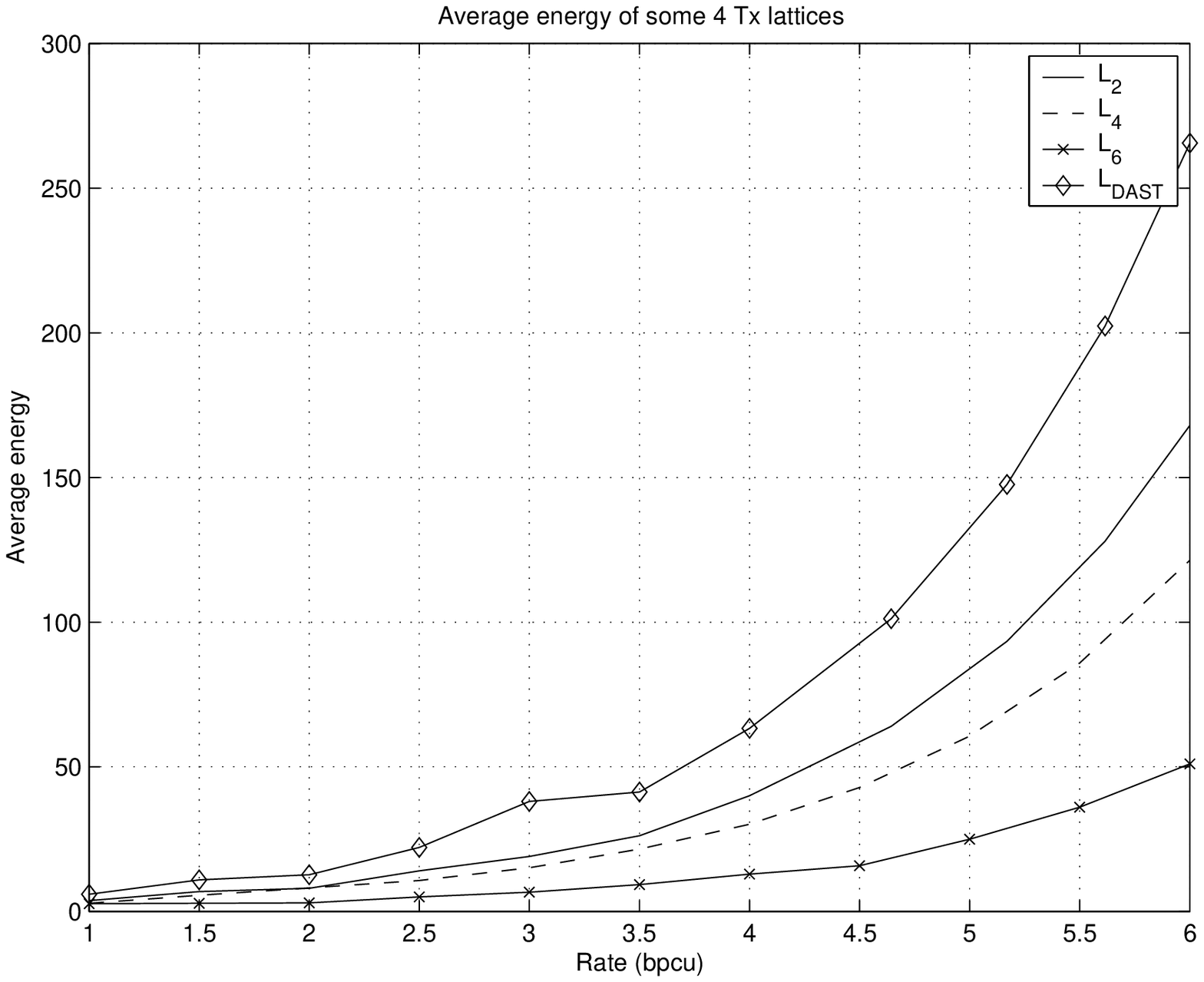}
\includegraphics[height=7.2cm]{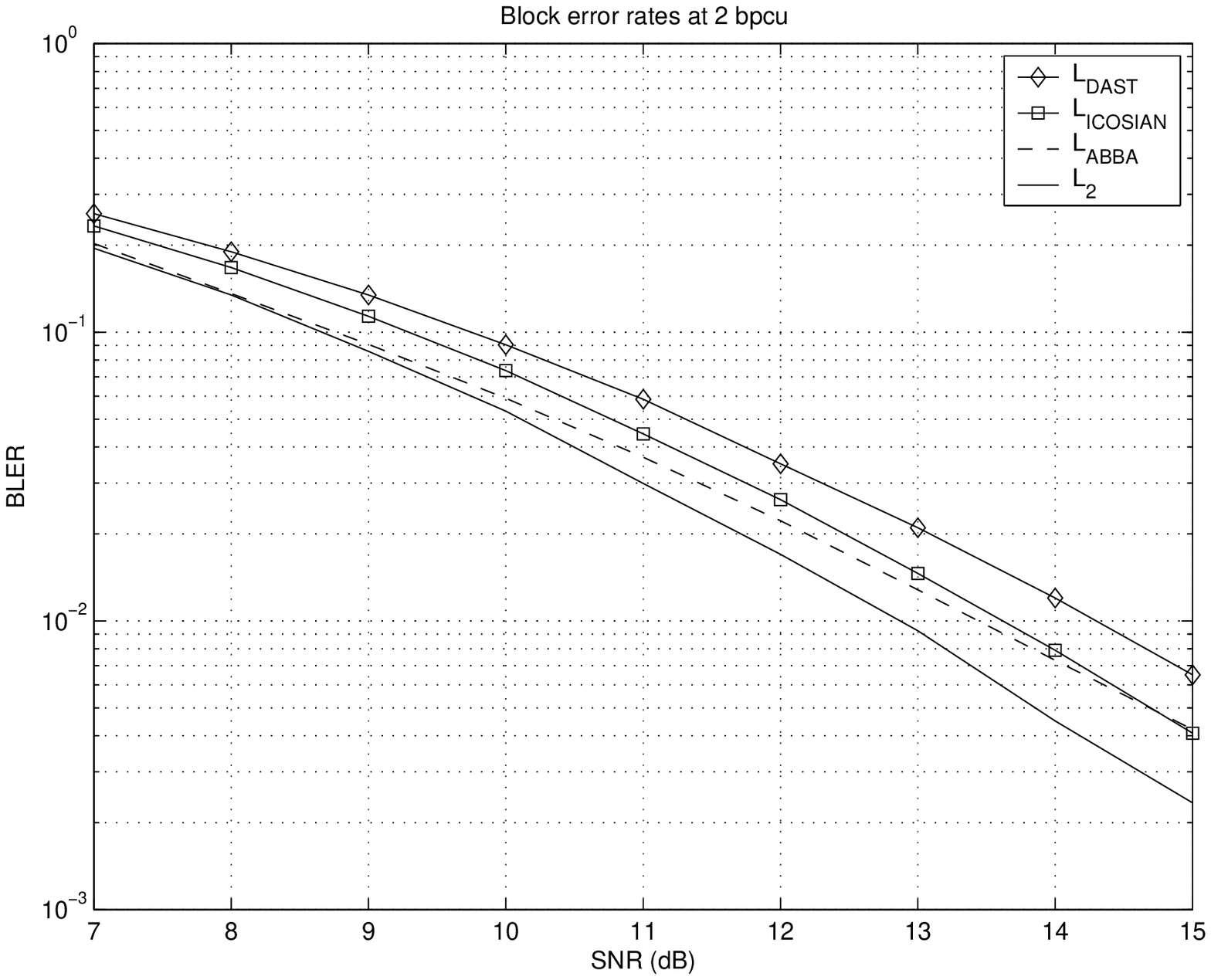}
\caption{Average energy (left) and block error rates of $4$ Tx-antenna lattices at $2$ bpcu
with one receiver (right).}\label{fig5}
\end{center}
\end{figure}

The exact average transmission power data in Fig. \ref{fig5} is computed as follows. Given the size $K$ of the code we
choose a random set of $K$ shortest vectors from each lattice. The average energy of the code $$
E_{av}=\frac{\sum_{x\in\mathcal{C}}\|x\|^2}{K} $$ is then computed
with the aid of theta functions \cite{CS}.  All the lattices
were normalized to have minimum determinant equal to 1.
When using the matrices $M(c_1,c_2,c_3,c_4)$ of Proposition \ref{algebras}, in some cases we are better off selecting the input vectors
$(c_1,c_2,c_3,c_4)$ from the coset
$\frac{1}{2}(1+i,1+i,1+i,1+i)+\G^4$
instead of letting them range over $\G^4$. Obviously such a translation
does not change the minimum determinant of the code, but it sometimes
results in significant energy savings. E.g. to get a code of size
256 it is clearly desirable to let the coefficients
$c_1,c_2,c_3,c_4$ range over the QPSK-alphabet.


\begin{figure*}[!ht]
\begin{center}
\includegraphics[height=7.3cm]{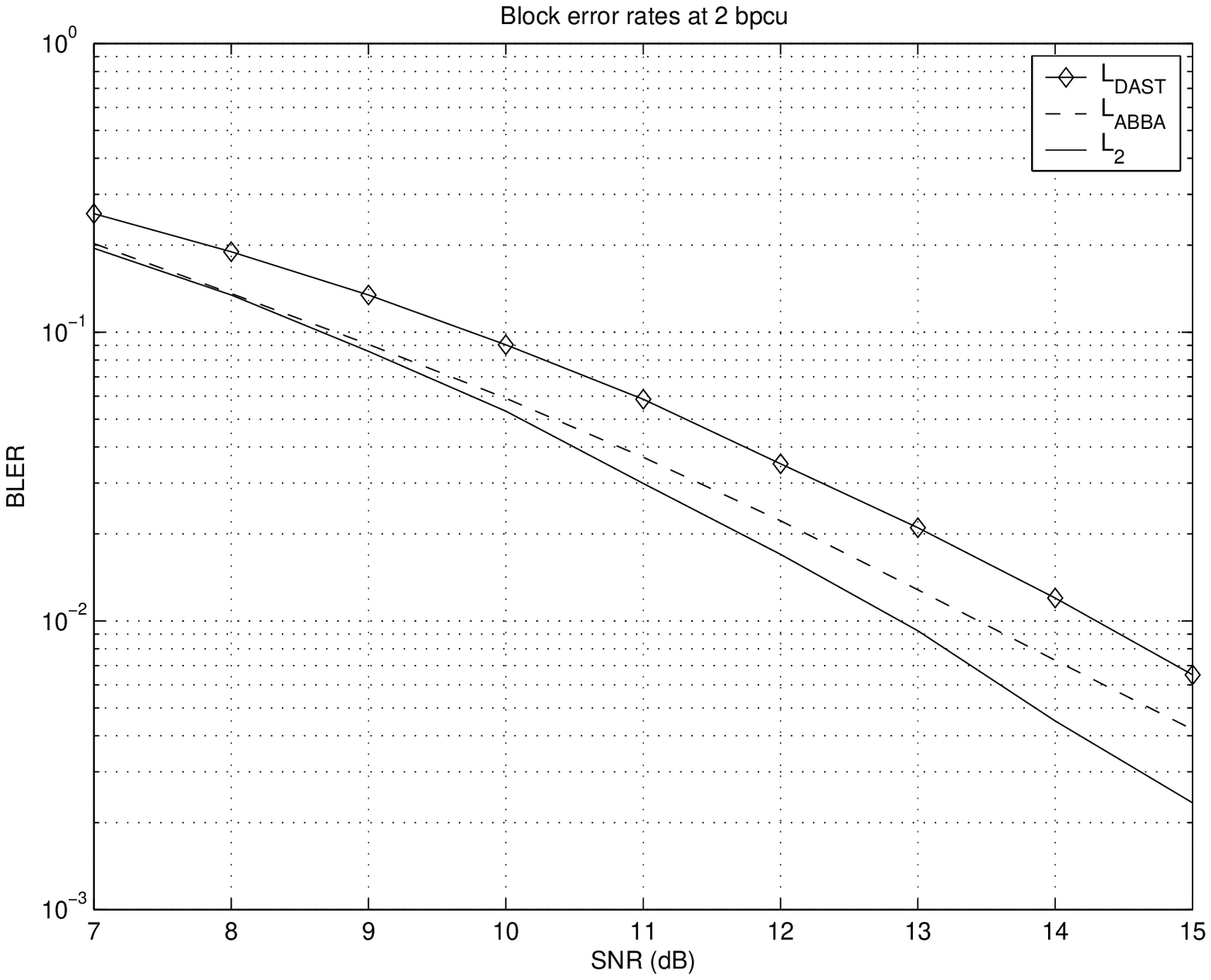}
\includegraphics[height=7.3cm]{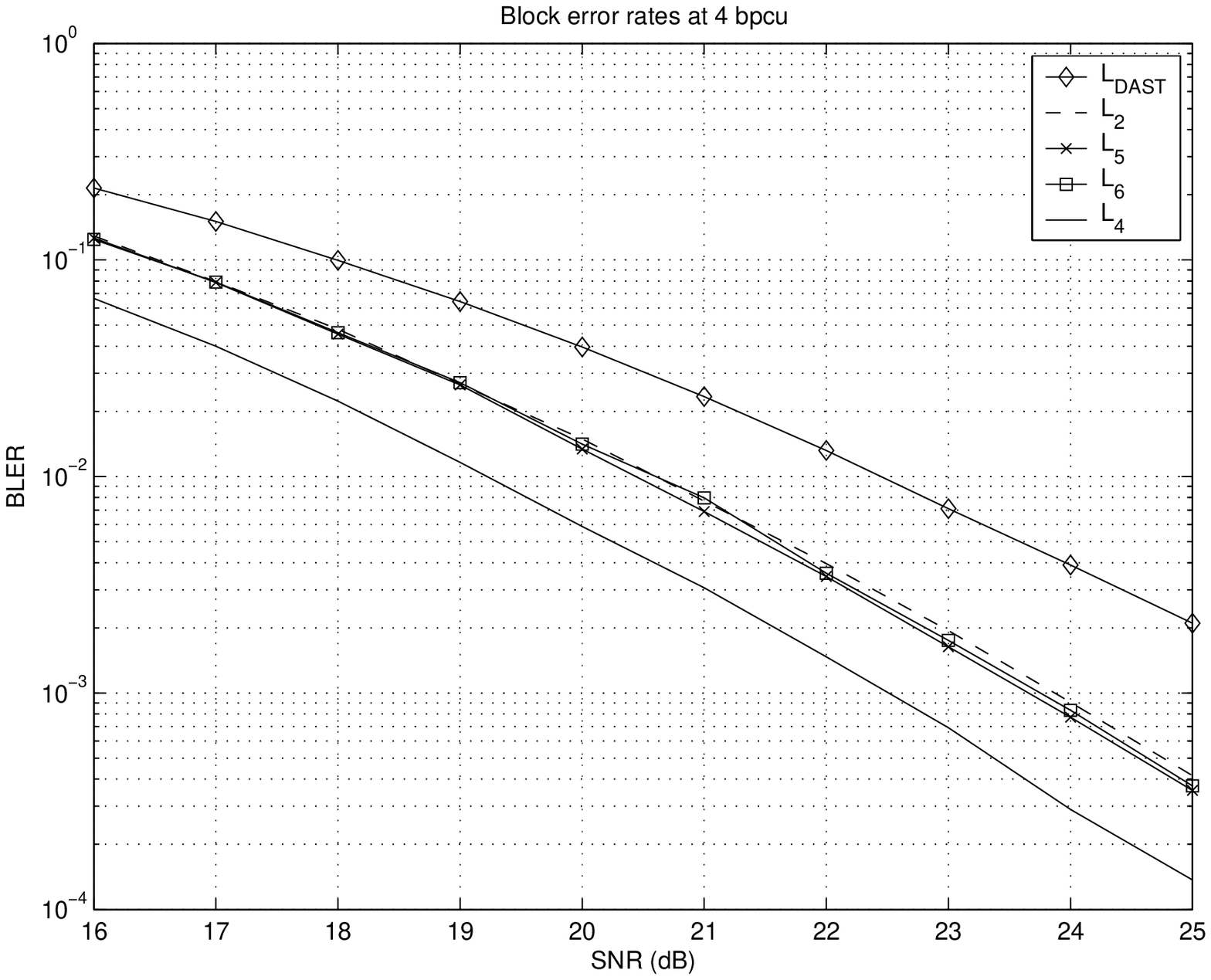}
\includegraphics[height=7.3cm]{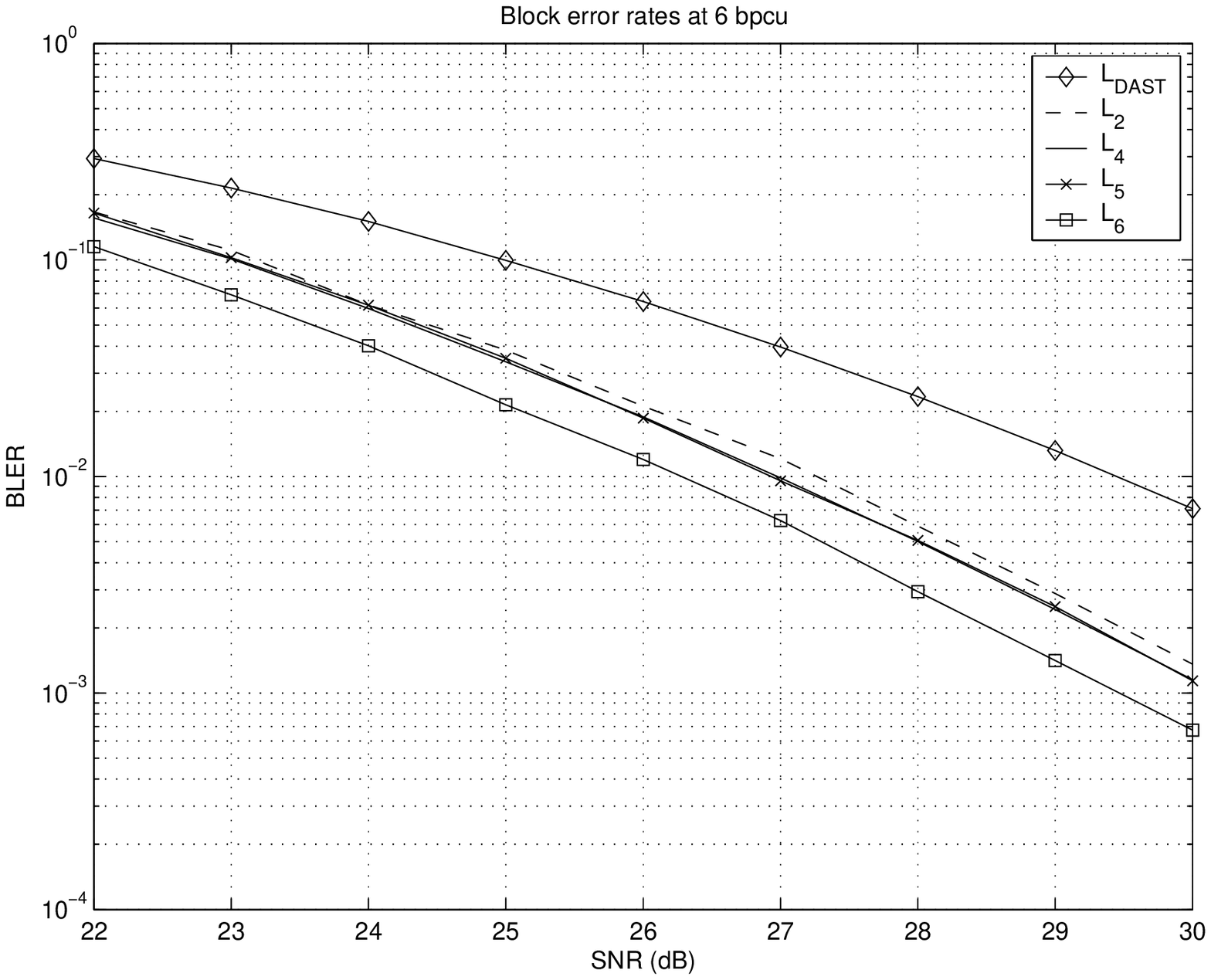}
\includegraphics[height=7.3cm]{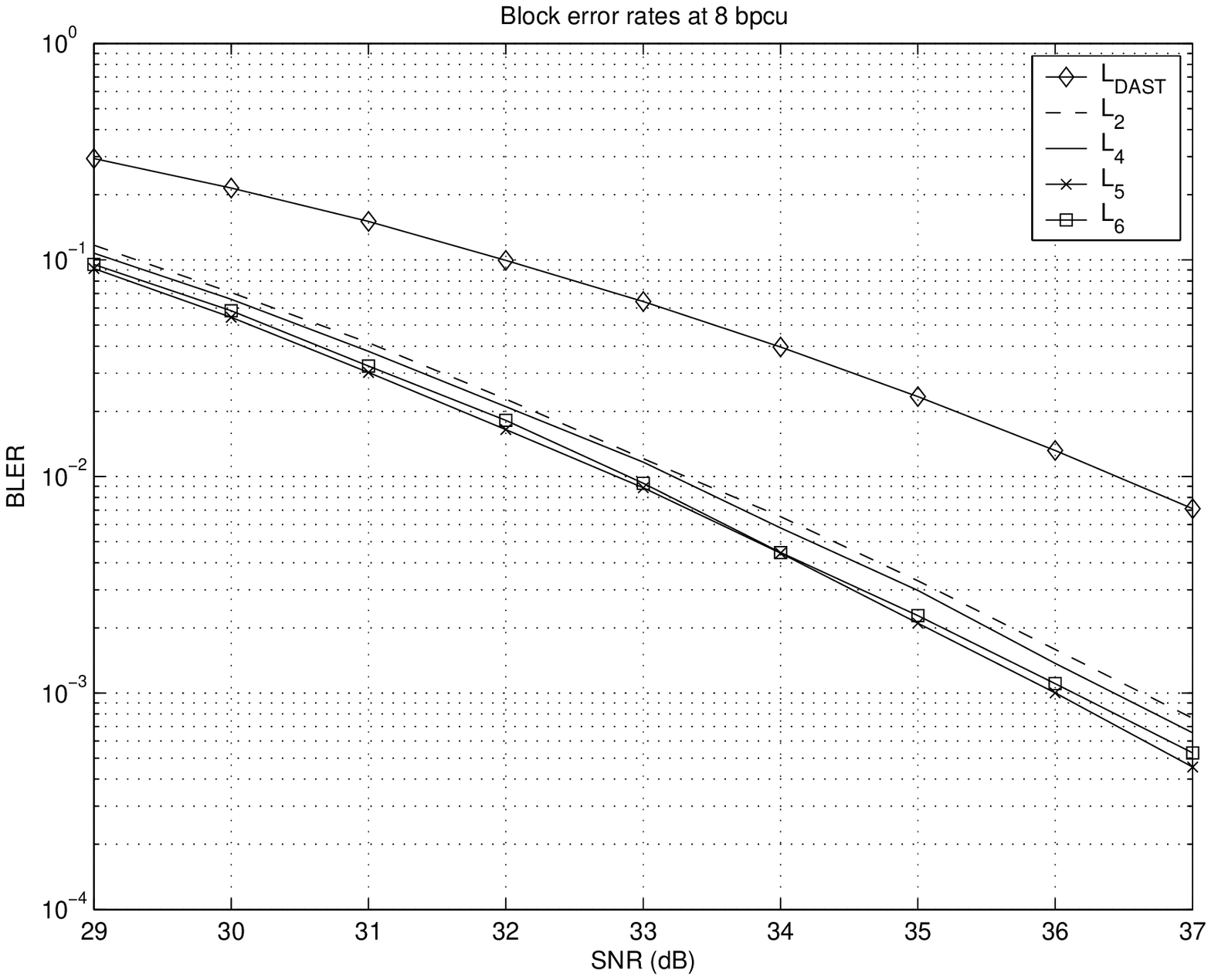}
\caption{Block error rates of $4$ tx-antenna lattices at approximately $2.0$, $4.0$, $6.0$, and $8.0$
 bpcu with one receiver.}\label{fig4}
\end{center}
\end{figure*}

Fig. \ref{fig4} shows the block error rates of the various
competing lattice codes at the rates approximately 2, 4, 6, and 8 bpcu, i.e. all the
codes contain roughly $2^8, 2^{16}, 2^{24}$ or $2^{32}$ matrices respectively. For the lattices $L_1$, $L_2$,
$L_{DAST}$, and $L_{ABBA}$ \cite{TBH} this simply amounted to
letting  the coefficients $c_1,c_2,c_3,c_4$ take all the values in
a QPSK-alphabet. Therefore, it would have been easy to obtain
bit error rates as well. For the lattices $L_4$, $L_5$,  $L_6$ the rate is not exact, see (\ref{subrate}) below and the preceding explanation. Of course also the exact rate equal to a  power of two could be achieved by just choosing a
more or less random set of  shortest lattice vectors. As
there is no natural way to assign bit patterns to vectors of
$D_8$, $D_4\bot D_4$ or $E_8$, we chose to show the block error
rates instead of the bit error rates.

The simulations were set up, here, so that the 95
per cent reliability range amounts to a relative error of about 3
per cent at the low SNR end and to about 10 per cent at the high
SNR end (or to about 4000 and 400 error events respectively).
One receiver was used for all the lattices.

When moving left in (\ref{nested
sequence}) the minimum determinant increaces while the BLER
decreases at the same time.
However,  the other side of the coin is that improvements
in the BLER performance cause a slightly more complex decoding
process by increasing the number of points visited in the search
tree. Still after this increasement, even the lattice $L_6$ admits
a fairly low average complexity as compared to the lattices $L_1$ and $L_{DAST}$ due to its lower sensitivity.
In part of the pictures in Fig. \ref{fig4}, the order of the curves seems not to respect the above mentioned order,
but this only happens because the rates are not exactly the same for all the lattices. E.g. at the rate
$\approx 4$ bpcu, the exact rates for $L_2,L_4,L_5,$ and $L_6$ are $4, 3.75, 4.14,$ and $ 4.17$ bpcu respectively.
Consequently, the lattice $L_4$ seems to perform better than what it actually does. Let us shortly explain
how these rates follow: when picking the elements $x_1,...,x_8$ from the set $\Z_Q$ (cf. Section \ref{decoding} (\ref{detector}) and the discussion after Algorithm II), the size of the code within the lattice $L_i,\ i=2,4,5,6$, will be $\frac{Q^8}{[L_2:L_i]}=2^{\textrm{log}\frac{Q^8}{[L_2:L_i]}}$, where $[L_2:L_i]$ is the index of the sublattice $L_i$ inside $L_2$ (cf. Proposition \ref{pdmin}). Hence, the data rate in bits per channel use can be computed as
\begin{equation}
\label{subrate}
R=\frac{\textrm{log}\frac{Q^8}{[L_2:L_i]}}{4}.
\end{equation}
Now, for instance, to get as close to the rate $R=4$ bpcu as possible, we have to choose $Q=4,Q=4,Q=5,$ and $Q=6$ for the lattices $L_2,L_4,L_5,$ and $L_6$ respectively. By substituting $Q$ and the sublattice index in question to (\ref{subrate})  we obtain the above rates.

Simulations at the rate $6$ bpcu
with one receiver show that the lattice $L_6$ wins by approximately $1$ dB over the lattice $L_2$ and by at least $2.5$ dB over $L_{DAST}$. At the rate $2$ bpcu, the rotated ABBA lattice $L_{ABBA}$ is already beaten by the $L_2$ lattice by a fraction of a dB. The difference between $L_2$ and $L_{DAST}$ is even clearer: $L_2$ gains $1-2$ dB over $L_{DAST}$, depending on the SNR. At all  data rates the lattice $L_6$ outperforms all the other lattices.


Prompted by the question of one of the reviewers, we make the following remark in case that the reader is familiar with the Icosian code  \cite{LC} and ponders over whether and how it relates to the codes presented in this paper.
\vspace*{4pt}
\begin{remark}
\label{icosianlattice}
The Icosian lattice $L_{ICOSIAN}$ presented in \cite{LC} takes use of the Icosian ring (cf. Remark \ref{icosian}) and has a similar looking structure to the Golden code \cite{BORV}, where the matrix elements are replaced with Icosian Alamouti blocks 
$$A=A(a_1,a_2,a_3,a_4)=\begin{pmatrix} a_1+a_2i&-a_3+a_4i\\a_3+a_4i&a_1-a_2i\end{pmatrix}$$ 
and $B=B(b_1,b_2,b_3,b_4)$ respectively:
$$L_{ICOSIAN}=\left\{\begin{pmatrix} A & K\overline{B}\\ B & \overline{A}\end{pmatrix} \ \Big |\   a_i,b_i\in\Z[(1+\sqrt{5})/2]\  \forall i \right\},$$
where $\overline{A}$ denotes the algebraic conjugate of $A$ with respect to the mapping $\sqrt{5}\mapsto -\sqrt{5}$ and $$K=\begin{pmatrix} i&0\\0&-i\end{pmatrix}.$$ A code within this lattice is called {\it Icosian code}. Note that Jafarkhani's quasi-orthogonal code \cite{J} in the simulations of \cite{LC} is exactly our base lattice $L_2$.

First of all, note that the Icosian code has code rate two, as the lattice is 16-dimensional over the reals. Hence, in order to enable efficient linear decoding, at least two antennas are required at the receiving end. Taking this into consideration, there is no good way to make fair comparison between the Icosian lattice and the 8-dimensional lattices proposed in this paper. If the application at hand allows us to use one receiving antenna only, we either have to puncture $L_{ICOSIAN}$ (e.g. by setting $B=0$) which will cause it to lose its benefits, or, we need to perform complex decoding process (e.g. a sphere decoder cannot be used). 

However, if we still want to compare these codes with two receivers, our codes will of course lose due to the lower code rate as they are designed for MISO use only. Similar comparison could be done e.g. with the $4\times4$ Perfect code \cite{BORV} and the Icosian code resulting to the loss of the Icosian code due to its lower rate (two vs. four). When using one receiver for the Icosian code by punctring the block $B$, it will lose to $L_2$ by 0.5-1 dB at 2 bpcu depending on the SNR as depicted in Figure \ref{fig5}.  But, as noted above, in this way $L_{ICOSIAN}$ will of course lose its benefits (as we are not really using the whole Icosian ring) so this is not a comparison on which we should put too much value.

To conclude, the codes in this paper and the Icosian code are targeted into different types of applications: the first ones are aimed for systems with one receiving antenna, whereas the Icosian code naturally fits into systems with two receiving antennas.
\end{remark}

\section{Diversity-multiplexing tradeoff analysis}
\label{DMT}
\newcommand{\tr}{\textrm{Tr}}
\newcommand{\polpl}{\ensuremath{S_{1,n-1}^{(n)} } }
\newcommand{\e}{\mathbf{e}}
\newcommand{\etal}{{\it et al. }}
\newcommand{\qed}{\hfill\IEEEQED}
\newcommand{\beq}{\begin{equation}}
\newcommand{\eeq}{\end{equation}}
\newcommand{\beqn}{\[}
\newcommand{\eeqn}{\]}
\newcommand{\bea}{\begin{eqnarray}}
\newcommand{\eea}{\end{eqnarray}}
\newcommand{\bean}{\begin{eqnarray*}}
\newcommand{\eean}{\end{eqnarray*}}
\newcommand{\re}{\mbox{$\mathfrak{Re}$}}
\newcommand{\bit}{\begin{itemize}}
\newcommand{\eit}{\end{itemize}}
\newcommand{\no}{\nonumber}
\newcommand{\ben}{\begin{enumerate}}
\newcommand{\een}{\end{enumerate}}
\newcommand{\dham}{d_{\text{H}}}
\renewcommand{\frak}{\mathfrak}
\newcommand{\In}{\textnormal{In}}
\newcommand{\disc}{\textnormal{disc}}
\newcommand{\Out}{\textnormal{Out}}
\newcommand{\cut}{\textnormal{cut}}
\renewcommand{\O}{\cal O}
\newcommand{\M}{\cal M}
\renewcommand{\_}{\underline}
\newcommand{\mf}[1]{{\textnormal{maxflow}}(#1)}
\renewcommand{\span}[1]{\left\langle #1 \right\rangle}
\newcommand{\<}{\left\langle}
\renewcommand{\>}{\right\rangle}
\newcommand{\rank}{\text{rank}}
\newcommand{\SNR}{\mbox{\textsf{SNR}}}
\newcommand{\now}{{\the\hour}:{\the\minute}, \today }
\newcount\minute    
\newcount\hour      
\newcount\hourMins  
\minute=\time    
\hour=\time \divide \hour by 60 
\hourMins=\hour \multiply\hourMins by 60
\advance\minute by -\hourMins 

This section contains the DMT analysis of the MISO codes constructed
in this paper. We denote by $n_t$ (resp. $n_r$) the number of transmitting (resp. receiving) antennas.  For the rest of the notation, see \cite{ZT}. 

Let us first consider the number field construction. Denote (cf. Proposition \ref{rings})
\[
L_1 \ = \ \left\{\left(
\begin{array}{cccc}
c_1 &  i c_4 & i c_3 & i c_2\\
c_2 & c_1 & i c_4 & i c_3\\
c_3 & c_2 & c_1 & i c_4\\
c_4 & c_3 & c_2 & c_1
\end{array} \right),  c_i \in {\cal A} \right\},
\]
where ${\cal A} \subset \Z [i]$ is some constellation set. This
code is for the MISO system with $n_t=4$ transmit and $n_r = 1$
receive antennas. Given the transmit code matrix $X \in L_1$,
the received signal vector is
\[
\underline{y}^T \ = \ \theta \underline{h}^T X + \underline{n}^T,
\]
where $\underline{h} \sim \C {\cal N}(\underline{0}, I_4)$.

Let $r$ be the desired multiplexing gain; then we need
\[
\left| L_1 \right| \ \doteq \ \SNR^{4r} \ \doteq \ \left| {\cal A} \right|^4
\]
and the above in turn gives
\beq
\left| {\cal A} \right| \ \doteq \ \SNR^r.
\eeq
Hence we see for every $c_i \in {\cal A}$
\beq
\left\| c_i \right\|^2 \ \dot\leq \ \SNR^r
\eeq
and
\beq
\theta^2 \ \doteq \ \SNR^{1-r}.
\eeq
Let $\lambda := \left\| \underline{h} \right\|_F^2 = \SNR^{-\alpha}$
and let $\delta_1 \geq \cdots \geq \delta_4$ be the ordered
eigenvalues of $XX^\dag$; then the random Euclidean distance $d_E$
is lower bounded by
\beq
d_E^2 \geq  \theta^2 \lambda \delta_4 \doteq
\frac{\theta^2\lambda}{\prod_{i=1}^3 \delta_i} \ \dot\geq \
\SNR^{E_{L_1}}
\eeq
where
\beq
{E_{L_1}} \ = \ 1 - r - \alpha - 3 r = 1 - 4 r - \alpha.
\eeq
Now the DMT of this code is given by
\beq
d_{L_1}(r) \ \geq \ \inf_{{E_{L_1}} < 0} 4 \alpha \ = \ 4 (1-4r), \ \ \text{
for $0 \leq r \leq \frac{1}{4}$,}
\eeq
while the optimal tradeoff in this channel is actually
\beq
d^*(r) \ = \ 4(1-r) \ \ \text{ for $0 \leq r \leq 1$.}
\eeq

The quaternionic construction is
\[
 L_2 \ = \ \left\{ \left(
\begin{array}{cccc}
c_1 &  i c_2 & -c_3^* & -c_4^*\\
c_2 & c_1 & i c_4^* & -c_3^*\\
c_3 & i c_4 & c_1^* & c_2^*\\
c_4 & c_3 & - i c_2^* & c_1^*
\end{array} \right), c_i \in {\cal A} \right\}.
\]
First of all, as pointed out in the proof of Proposition 2.4, the matrix $M \in
L_2$ is of the following form:
\[
M \ = \ \left(
\begin{array}{cc}
A & - B^H\\
B & A^H
\end{array} \right)
\]
and
\begin{eqnarray*}
M M^H \ &=& \ \left(
\begin{array}{cc}
AA^H + B^H B & {\bf 0}\\
{\bf 0} & A^HA + B B^H
\end{array} \right)\\ &=& \left(
\begin{array}{cc}
AA^H + B B^H & {\bf 0}\\
{\bf 0} & AA^H + B B^H
\end{array} \right)
\end{eqnarray*}
since $AB=BA$. Thus the ordered eigenvalues of $MM^H$ satisfy
$\delta_1 = \delta_2 \geq \delta_3 = \delta_4$ and in particular,
$\delta_1 \geq \delta_3$ are the ordered eigenvalues of $AA^H + B
B^H$. Secondly, note that $MM^H$ satisfies the non-vanishing
determinant property, and so does the matrix $AA^H + B B^H$. Now the
bound for the random Euclidean distance is
\beq
d_E^2 \ \geq \ \theta^2 \lambda \delta_4 \ \doteq \ \frac{\theta^2
\lambda}{\delta_3} \ \dot\geq \SNR^{E_{L_2}},
\eeq
where
\begin{equation}
E_{L_2} \ = \ 1 - r - \alpha - r = 1 - 2 r - \alpha.
\eeq
Now the DMT of this code is given by
\begin{equation}
d_{L_2}(r) \ \geq \ \inf_{{E_{L_2}} < 0} 4 \alpha \ = \ 4 (1-2r), \ \ \text{
for $0 \leq r \leq \frac{1}{2}$.} 
\eeq
The same of course also holds for codes within the sublattices $L_4,L_5,L_6\subseteq L_2$.
\vspace*{4pt}
\begin{remark}
While our codes are not DMT optimal, it has to be noticed that without using a full-rate code the DMT cannot be achieved. Hence, if one wishes to enable efficient decoding process with one receiving antenna only (see the remark below), sacrifices in terms of the DMT have to be made. However, our quaternionic lattices $L_2,L_4,L_5,L_6$ admit higher DMT as e.g. the DAST lattice, as the DMT of the DAST lattice coincides with that of $L_1$.
\end{remark}
\vspace*{4pt}
\begin{remark} One might ponder why not use e.g. the full-rate CDA based codes (cf. \cite{SRS}, \cite{BORV}) as they are DMT optimal provided that they have non-vanishing determinant. The answer to this is in principle the same as the one provided in Remark \ref{icosianlattice}.  We  could naturally do this, but considering that we only want to use one receiving antenna it should be clear that a full-rate code cannot be efficiently used. Indeed, using a full-rate code  would destroy the lattice structure and cause exponential complexity at the receiver. To enable efficient decoding with one receiver we have to limit ourselves to rate-one codes, which exactly we have done in this paper. We want the reader to note that full-rate codes (e.g. the perfect codes \cite{BORV}) are optimally suited  for systems with  $n_t=n_r>1$, hence inapplicable to the purposes of this paper where we have  $n_t=4$ and $n_r=1$.

\end{remark}

\bibliography{IEEEabrv,space,book}

\section{Conclusions and suggestions for further research}
\label{conclusions}

In this paper, we have presented new constructions of rate-one, full-diversity, and energy efficient $4\times 4$ space-time codes with non-vanishing determinant by
using the theory of rings of algebraic integers and their
counterparts within the division rings of Lipschitz' and Hurwitz'
integral quaternions. A comfortable, purely number theoretic way to
improve space-time lattice constellations was introduced. The use
of ideals provided us with denser lattices and an easy way to present
the exact proofs for the minimum determinants. The constructions can
be  extended also to a larger number of transmit antennas, and
they nicely fit with the popular Q$^2$-QAM and QPSK modulation alphabets.
The idea of finding denser sublattices within a given division algebra was also generalized to a MIMO case with arbitrary number of Tx antennas by using the theory of cyclic division algebras  and, as a novel method, their maximal orders. This is encouraging as the CDA based square ST constructions with NVD are known to achieve the DMT. We  have also shown that the explicit constructions in this paper all have a simple decoding method based on sphere decoding. Related to the decoding complexity,
the notion of sensitivity was introduced for the first time in this paper. The experimental results have given
evidence about the relevance of this new notion.

Comparisons with the four antenna DAST block code have shown that our codes provide lower energy
and block error rates due to their good minimum determinant, i.e.
high density and lower sensitivity. At the moment, we are searching for well-performing MIMO codes
arising from the theory of crossed product algebras and maximal orders
of cyclic division algebras. We have noticed that also the discriminant of a maximal order plays an important role in code design. It is desirable to choose cyclic division algebras for which the discriminant of a maximal order is as small as possible \cite{HLRV}. By now, we are able to  construct an explicit cyclic division algebra  of an arbitrary index over $\Q(i)$ (or $\Q(\omega)$) that has a maximal order with minimal discriminant. Despite the fact that we have not yet fully analyzed the practical performance of codes arising from these constructions, the preliminary results have been very promising.
 Further details on this and on the algorithmic properties of maximal orders (see also \cite{Ron}-\cite{M}) will be given in a forthcoming paper \cite{HLRV}.

\section{Acknowledgments} The authors are grateful to graduate student Miia Mäki for partly implementing the sphere decoder that was used for the simulations. A thank-you is also due to anonymous reviewers for their insightful comments that greatly improved  the quality of this paper.

C. Hollanti was supported in part by the Nokia Foundation, the Foundation of
Technical Development, and the Foundation of the Rolf Nevanlinna
Institute, Finland.


\end{document}